\documentclass[pdflatex,sn-mathphys-num]{sn-jnl}
\usepackage{graphicx}%
\usepackage{multirow}%
\usepackage{amsmath,amssymb,amsfonts}%
\usepackage{amsthm}%
\usepackage{mathrsfs}%
\usepackage[title]{appendix}%
\usepackage{xcolor}%
\usepackage{textcomp}%
\usepackage{manyfoot}%
\usepackage{booktabs}%
\usepackage{algorithm}%
\usepackage{algorithmicx}%
\usepackage{algpseudocode}%
\usepackage{listings}%
\theoremstyle{thmstyleone}%

\theoremstyle{thmstyletwo}%

\theoremstyle{thmstylethree}%

\usepackage{ulem}

\begin{document}

\title[Cosmological perturbations in Energy-Momentum Squared Gravity]{Cosmological perturbations in Energy-Momentum Squared Gravity}

\author[1,2,3]{\fnm{Peter K. S.} \sur{Dunsby}}\email{peter.dunsby@uct.ac.za}

\author[1]{\fnm{Maria-Alexia} \sur{Caldis}}\email{CLDMAR008@myuct.ac.za}

\author*[4]{\fnm{Eduardo} \sur{Bittencourt}}\email{bittencourt@unifei.edu.br}

\affil[1]{Department of Mathematics and Applied Mathematics, University of Cape Town, Rondebosch 7700, Cape Town, South Africa}

\affil[2]{Center for Space Research, North-West University, Potchefstroom 2520, South Africa}

\affil[3]{South African Astronomical Observatory, Observatory 7925, Cape Town, South Africa}

\affil[4]{Federal University of Itajub\'a, Av. BPS, 1303, Pinheirinho, Itajub\'a/MG - Brazil}

\abstract{We present a fully covariant and gauge–invariant analysis of linear cosmological perturbations in Energy-Momentum Squared Gravity. Working within the $1\!+\!3$ formalism, we derive the exact propagation equations for scalar, vector, and tensor modes on FLRW backgrounds, in the case of radiation and dust. Two representative subclasses are examined in detail, in which non-linearity enters through $\mathcal{O}(\eta\rho^2)$ corrections or modifications in the equation-of-state parameter and the sound speed. For scalar perturbations, the density contrast can be enhanced or reduced relative to General Relativity, depending on the coupling parameter and the wavelength regime. A similar behavior occurs for vector modes, allowing for a non-trivial vorticity at early times. Tensor modes, described by the magnetic part of the Weyl tensor and the shear tensor propagate as damped waves with slowly varying effective masses. All sectors reduce continuously to their GR limits as $\eta\!\to\!0$. The framework isolates robust signatures---early-time scalar tilts, tensor damping shifts, and altered vorticity decay---that can be confronted with CMB and large-scale-structure observations to constrain these theories of gravity.}


\maketitle

\section{Introduction}
The $\Lambda$CDM cosmological model, based on General Relativity (GR), together with cold dark matter and a cosmological constant, has been remarkably successful in accounting for cosmological observations from the cosmic microwave background to large-scale structure \cite{Weinberg2008,HawkingEllis1973}. Nonetheless, both conceptual and observational puzzles — including the nature of dark energy and dark matter, the status of cosmological singularities, and possible signatures of high-energy corrections to gravity — motivate the exploration of modified theories of gravity and of non-standard matter couplings (see, e.g., \cite{Clifton2012,DeFelice:2010aj,Harko2011,Joyce:2014kja} and references therein). Such matter-sector modifications are conceptually related to a broader class of cosmological models in which the dark components interact or exchange energy-momentum, as in coupled quintessence and interacting dark-sector scenarios~\cite{Amendola:1999er,Wang:2016lxa}. Recent large-scale-structure and baryon-acoustic-oscillation measurements, including DESI results, have also renewed interest in testing departures from the minimal \(\Lambda\)CDM framework and in constraining beyond-\(\Lambda\)CDM scenarios \cite{DESI:2024mwx,DESI:2025zgx}.

A closely related but distinct class of matter-coupled modified gravities arises when the gravitational Lagrangian is taken to depend explicitly on matter scalars constructed from the energy–momentum tensor. Two common choices in the literature are functions of the trace, $f(R,T)$, and functions of self-contraction, $f(R,\mathcal{T})$ with $\mathcal{T}\equiv T_{\mu\nu}T^{\mu\nu}$. The former family, pioneered in \cite{Harko2011}, introduces a direct dependence on the trace $T=g^{\mu\nu}T_{\mu\nu}$ and typically produces non-minimal matter–geometry couplings, which can lead to non-conservation of the matter energy–momentum tensor and to an extra (non-geodesic) force acting on test particles; these features have motivated a large body of follow-up work and critical assessments \cite{Barrientos2014,Velten2017,NojiriOdintsov2011,Akarsu2018a,Akarsu2020}. By contrast, theories depending on $\mathcal{T}$ (often called Energy–Momentum Squared Gravity, EMSG) couple to quadratic combinations of the components of $T_{\mu\nu}$ and therefore predominantly modify the dynamics in high-density regimes, producing novel high-energy phenomenology such as effective pressure terms, modified maximal densities or bounce solutions in some branches of solutions \cite{Katirci2014,Roshan2016,BoardBarrow2017,Barbar2020,Khodadi2022,Cipriano2024}, constraining the nucleosynthesis \cite{Akarsu2024}, structure formation \cite{Kazemi2020,Farsi2023} and compact-object physics \cite{Akarsu2018b}.

Because these two families modify gravity in rather different ways, it is useful to adopt a clear terminology: in this work, we reserve the name ``EMSG'' for theories of the form $F\big(R,\;\mathcal{T}\big) = R + \eta\big(\mathcal{T}\big)^{n}$, $n\in\mathbb{R}$, so that EMSG denotes the whole class and the commonly studied cases $n=1$ (quadratic correction) and $n=\tfrac12$ (square-root scaling) are treated as representative sub-models. This choice is convenient because (i) the $\mathcal{T}$ coupling directly controls deviations that become important at high energy densities (early universe, compact objects), and (ii) the algebraic structure of the correction allows an ``effective fluid'' interpretation (effective $\bar\rho$, $\bar p$ and $\bar c_s^2$) that greatly simplifies the covariant perturbation analysis presented below. Nonetheless, both families have caveats: $f(R,T)$ models require careful handling of conservation laws and have been shown to face cosmological viability challenges in some parameterizations \cite{Barrientos2014,Velten2017}, while EMSG models must be scrutinized for consistent microphysical interpretation (choice of $\mathcal{L}_m$), stability and observational constraints at high densities \cite{Roshan2016,BoardBarrow2017}. On balance, using $F(R,\mathcal{T})$ as the study case is a good and transparent choice for scrutinizing high-density phenomenology and linear perturbations in a covariant, gauge-invariant way.

For the study of cosmological perturbations, a number of complementary formalisms exist. The metric-based gauge-invariant approach (Bardeen variables) \cite{Bardeen1980,KodamaSasaki1984,Mukhanov1992,MaBertschinger1995,Malik:2008im} and the covariant 1+3 approach \cite{StewartWalker1974,Ellis1989a,Ellis1989,Ellis1990,EllisLectures1991, Bruni:1992dg, Dunsby:1991xk, Tsagas2008} provide different but equivalent descriptions at linear order and are widely used in cosmology. In particular, the 1+3 covariant and gauge-invariant framework is particularly well suited to our goals because it employs physically transparent variables (such as the comoving fractional density gradient and expansion inhomogeneities), makes the role of kinematic quantities (expansion, shear, vorticity, acceleration) explicit, and avoids gauge ambiguities by construction. This approach also cleanly exposes couplings between scalar, vector and tensor sectors (for instance, the influence of vorticity on density gradients), is convenient for discussing long-wavelength limits, conserved quantities and invariant decompositions \cite{Ellis1989,Ellis1990,Dunsby1997}, and has already proved useful in the covariant
analysis of perturbations and dynamics in modified-gravity models, in particular in \(f(R)\) gravity \cite{JPhysA4025S40,PhysRevD83024030,CQG26235018}. It also admits a dynamical system and quantization approach
\cite{Novello1995a,Novello1995b,Novello1996,NOvello2014}.

In this paper, we extend the 1+3 covariant perturbation formalism to EMSG models aforementioned, specializing where useful to the commonly studied cases $n=1$ and $n=1/2$. Our main goals are: (i) to derive the linear evolution equations for covariant, gauge-invariant density perturbations in the effective-fluid interpretation of EMSG; (ii) to identify how the non-linear dependence of the theory on the matter
invariant \(T_{\mu\nu}T^{\mu\nu}\) modifies the effective background equation-of-state and sound-speed parameters that enter the linear perturbation equations; (iii) to include carefully the effects of vorticity and to present the invariant local decomposition of the density-gradient field into its irreducible parts following Ellis et al.\ \cite{Ellis1990}; (iv) to analyze the resulting generalized Jeans criterion and discuss qualitative consequences for structure formation in the early universe; and, finally, (v) to investigate the propagation of gravitational waves.

The structure of the paper is as follows. In Sec.\ II we present the EMSG action and field equations, obtain the homogeneous and isotropic background equations, and introduce the effective energy density and pressure. In Sec.\ III, we develop the linear perturbation equations in the 1+3 covariant formalism, derive the second-order propagation equation for the comoving fractional density gradient (including the vorticity source term), and discuss special cases. In Sec.\ IV, we present an invariant local decomposition of the density gradient, scrutinize the evolution equation for the scalar part, and perform a harmonic decomposition. We also discuss the generalized Jeans instability in EMSG and present analytic and numerical results for dust and radiation regimes for the models of interest. In Sec.\ V, we study the propagation equation of tensor modes in the form of primordial gravitational waves. We conclude in Sec.\ VI with a summary and possible observational/phenomenological implications. We work in geometric units with $c=1$ and $8\pi G=1$.

\section{Revisiting background cosmology in $F(R,\mathcal{T})$ gravity}
\label{sec:background}

In this section, we briefly review and extend well-known results of EMSG in a homogeneous and isotropic background. We introduce an effective-fluid interpretation, which will be used throughout the paper, and discuss special cases and physically interesting limits. 

\subsection{Action and field equations}

The class of models considered is driven by the action
\begin{equation}
\label{eq:action}
S = \frac{1}{2}\int d^4x\sqrt{-g}\,F\!\big(R,\,\mathcal{T}\big) - \int d^4x\sqrt{-g}\,\Lambda
    + \int d^4x\sqrt{-g}\,\mathcal{L}_m,
\end{equation}
where $\Lambda$ represents the cosmological constant and $\mathcal{L}_m$ is the Lagrangian for the matter content. The energy-momentum tensor is defined as 
$$T_{\mu\nu}=-\frac{2}{\sqrt{-g}}\frac{\delta(\sqrt{-g}\mathcal{L}_m)}{\delta g^{\mu\nu}}.$$ 
As we said, in this paper, we will focus on the family
\begin{equation}
\label{eq:F_choice}
F\big(R,\mathcal{T}\big) = R + \eta \big(\mathcal{T}\big)^n,
\qquad \eta=\text{const},\quad\mbox{and}\quad  n\in\mathbb{R}.
\end{equation}
In particular,  we will study the cases $n=1$ and $n=1/2$, hereafter referred to as \textit{Model A} and \textit{Model B}, respectively. Note that for each value of the exponent, the parameter $\eta$ has a different physical dimension; for instance, in Model A it has the dimension of the inverse of the energy density, whereas in Model B it is dimensionless. The variation of Eq.\ \eqref{eq:action} with respect to $g^{\mu\nu}$ gives the field equations (see Refs.~\cite{Roshan2016,BoardBarrow2017} for details):
\begin{equation}
\label{eq:field_eq_general}
G_{\mu\nu} + \Lambda g_{\mu\nu}
= T_{\mu\nu} + \eta\,(\mathcal{T})^{\,n-1}\!\left[\frac{1}{2}(\mathcal{T})g_{\mu\nu} - n\,\Theta_{\mu\nu}\right],
\end{equation}
where
\begin{equation}
\Theta_{\mu\nu} = -2\mathcal{L}_m\left(T_{\mu\nu}-\tfrac{1}{2}T g_{\mu\nu}\right) - T\,T_{\mu\nu} + 2T_{\mu}{}^{\alpha}T_{\alpha\nu} - 4 T^{\alpha\beta}\frac{\partial^2\mathcal{L}_m}{\partial g^{\mu\nu}\partial g^{\alpha\beta}}.
\end{equation}
The matter Lagrangian is taken to be that of a perfect fluid, following the common choice $\mathcal{L}_m=p$ (see discussion in \cite{Harko2011,Schutz1970}). 
With this assumption, the last term vanishes under the minimal coupling, and the form of $\Theta_{\mu\nu}$ simplifies\footnote{We emphasize that alternative action-based formulations of relativistic fluids \cite{SchutzSorkin:1977,Brown:1993,Dubovsky:2012,Akarsu2024eq} generally lead to a nontrivial metric dependence of $\mathcal{L}_m$, for which this second-derivative contribution need not vanish. Our paper applies to the conventional field-equation realization widely used and exploring the corresponding model-dependence of $\Theta_{\mu\nu}$ is beyond the scope of the present paper.}. Needless to say, setting $\eta\to 0$, GR with cosmological constant is recovered, and in vacuum ($T_{\mu\nu}=0$), the additional terms vanish and standard vacuum GR is obtained.

\subsection{The background metric and effective--fluid interpretation}
\label{sec:flrw}

For the background geometry, we assume a spatially homogeneous and isotropic Friedmann-Lema\^itre-Robertson-Walker (FLRW) metric
\begin{equation}
\label{eq:flrw_metric}
ds^2 = -dt^2 + a^2(t)\left[\frac{dr^2}{1-\epsilon r^2} + r^2(d\theta^2+\sin^2\theta\,d\varphi^2)\right],
\end{equation}
with $a(t)$ as the scale factor and $\epsilon$ as the spatial 3-curvature, which can be zero (flat), positive (closed), or negative (open). For the matter content, we consider a class of comoving normalized timelike vector fields $u^{\mu}=\delta^{\mu}_0$ decomposing the cosmic web as a perfect-fluid energy-momentum tensor
\begin{equation}
T_{\mu\nu}=\rho\,u_\mu u_\nu + p\,h_{\mu\nu},\qquad h_{\mu\nu}=g_{\mu\nu}+u_\mu u_\nu,
\end{equation}
where $\rho$ is the physical energy density, $p$ is the barotropic pressure, and $h_{\mu\nu}$ is the projector onto the 3-space orthogonal to $u^{\mu}$.

For the family \eqref{eq:F_choice}, the modified Friedmann and acceleration equations read \cite{BoardBarrow2017}
\begin{align}
\label{eq:friedmann_general}
&H^2 + \frac{\epsilon}{a^2} = \frac{\Lambda}{3} + \frac{\rho}{3} + \frac{\eta}{3}\,(\rho^2+3p^2)^{\,n-1}\Big[\Big(n-\frac{1}{2}\Big)(\rho^2+3p^2) + 4n\,\rho p\Big],\\[1ex]
\label{eq:acc_general}
&\frac{\ddot a}{a} = -\frac{1}{6}(\rho+3p) + \frac{\Lambda}{3} - \frac{\eta}{3}\,(\rho^2+3p^2)^{\,n-1}\Big[\frac{n+1}{2}(\rho^2+3p^2) + 2n\,\rho p\Big],
\end{align}
where $H\equiv\dot a/a$ is the Hubble parameter and the overdot denotes the derivative with respect to cosmic time. Note the non-linear contribution of the source to the background equations.

Because the extra terms are algebraic functions of $\rho$ and $p$, it is useful to cast the modifications as an \emph{effective fluid}:
\begin{align}
\bar{\rho} &\equiv \rho + \rho_{\rm emsg},\qquad
\rho_{\rm emsg} \equiv \eta(\rho^2 + 3p^2)^{n-1}\left[\left(n-\frac{1}{2}\right)(\rho^2 + 3p^2) + 4n\rho p\right],\label{eq:rho_eff}\\
\bar{p} &\equiv p + p_{\rm emsg},\qquad
p_{\rm emsg} \equiv \frac{\eta}{2}\big(\rho^2+3p^2\big)^n.\label{eq:p_eff}
\end{align}
With these definitions, Eqs.~\eqref{eq:friedmann_general}--\eqref{eq:acc_general} take the GR form with sources $\bar\rho$ and $\bar p$, as follows
\begin{equation}
    \label{eq:eff_Friedmann}
H^2 + \frac{\epsilon}{a^2} = \frac{\Lambda}{3} + \frac{\bar\rho}{3},\qquad
\frac{\ddot a}{a} = -\frac{1}{6}(\bar\rho+3\bar p) + \frac{\Lambda}{3}.
\end{equation}
Hence, one can reutilize many standard cosmological tools by replacing $(\rho,p)$ with $(\bar\rho,\bar p)$, including linear perturbation analysis, as we shall see. 

\begin{figure}
    \centering
    \includegraphics[width=0.9\linewidth]{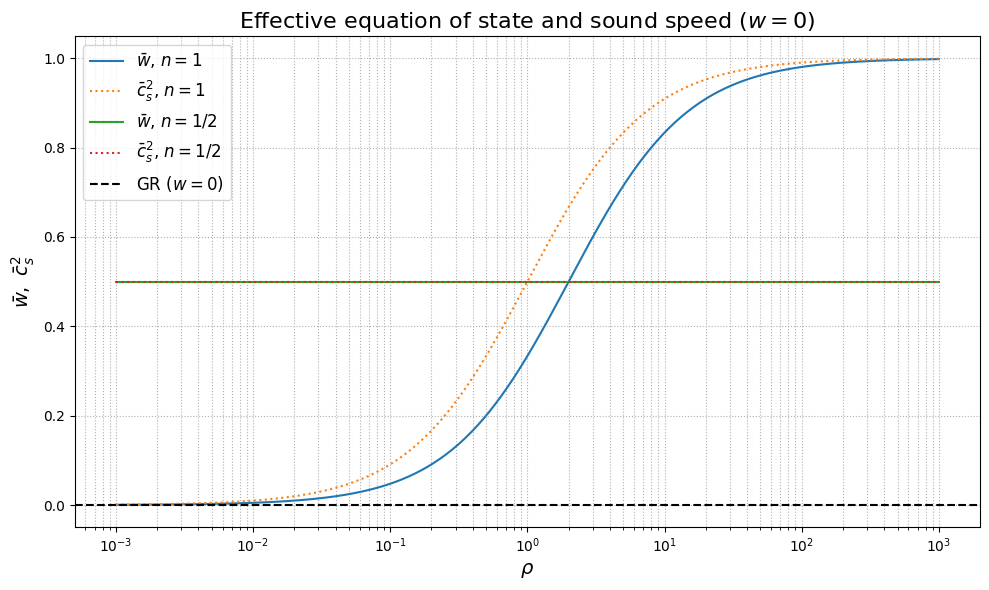}
    \caption{Effective equation-of-state parameter and sound speed for Models A and B, in the case of dust. For the sake of illustration, we set $\eta=1.0$ for both models.}
    \label{fig:eff_w_cs2_dust}
\end{figure}

For a barotropic equation of state $p=w\rho$, one obtains closed-form expressions for the effective equation of state parameter $\bar w$ and the effective adiabatic sound speed $\bar c_s$; that is,
\begin{align}
\label{eq:bar_w}
    \bar{w} &\equiv \frac{\bar{p}}{\bar{\rho}} = \frac{w + \frac{\eta}{2}(1 + 3w^2)^n\rho^{2n-1}}{1 + \eta A(n,w)\rho^{2n-1}},\\[1ex]
\label{eq:bar_cs2}
    \bar{c}_s^2& \equiv \frac{d\bar p}{d\rho}\frac{d\rho}{d\bar\rho}=\frac{w + n\,\eta\,\rho^{2n-1}\,(1 + 3w^2)^{n}}{1 + 2n\,\eta\, \rho^{2n-1}\,A}.
\end{align}
where $A(n,w)=(1 + 3w^2)^{n-1}[(2n-1)(1 + 3w^2) + 8nw]/2$. Remarkably, for radiation ($w=1/3$) in Model A ($n=1$), one obtains $\bar w=\bar c_s^2=1/3$, which means that the physical and effective fluids possess the same parameters, although the physical and effective variables behave distinctly. A similar behavior occurs in Model B ($n=1/2$), where $\bar w=\bar c_s^2=\mbox{const.}$, but $\bar w>w$ for positive small $\eta$'s and $\bar w<w$ for negative small $\eta$'s. For incoherent matter ($w=0$), for instance, an effective pressure and sound speed appear:
\begin{equation}
\label{eq:dust_eff}
\bar{w}=\frac{\eta\rho^{2n-1}}{2+(2n-1)\eta\rho^{2n-1}},\qquad \bar{c}_s^2=\frac{n\,\eta\,\rho^{2n-1}}{1+(2n-1)n\,\eta\,\rho^{2n-1}},
\end{equation}
illustrating that EMSG generically endows dust with effective pressure, which will have a significant impact on the evolution of perturbations (as we will see later). In fact, from Fig.\ (\ref{fig:eff_w_cs2_dust}), it is possible to see that for small matter densities, the corrections are perturbations around the GR limit, and at high densities, the effective fluid tends to stiff matter ($w\rightarrow1$), which can lead to a modified early-universe expansion or even a bounce. From Fig. (\ref{fig:eff_w_cs2_dust}), together with Eqs. (\ref{eq:bar_w}) and (\ref{eq:bar_cs2}), one can also see that for $\eta>0$ the models avoid instabilities (ghost or tachyonic modes), thus ensuring the perturbative analysis remains well behaved.

\subsection{Continuity equation and integrability}
\label{sec:continuity}

Conservation of the (total) energy-momentum tensor of matter is still implied since the extra terms modify the field equations but do not break diffeomorphism invariance. Hence, the effective energy-momentum tensor obeys $\nabla^\mu \bar T_{\mu\nu}=0$. Working on the FLRW background, for the general power $n$ in Eq.\ \eqref{eq:F_choice}, the dynamics entirely in terms of $\rho$ can be written in the form (cf. details in Ref.\ \cite{BoardBarrow2017})
\begin{equation}
\left[\frac{1 + 2n\,\eta\,\rho^{2n-1}A}{1+n\,\eta\,\rho^{2n-1}(1+3w^2)^{n-1}(1+3w)}\right]\frac{\dot\rho}{\rho}=-3H\,(1 + w).
\end{equation}
By a suitable choice of variable, this equation can be integrated for arbitrary $n$ and $w$, yielding
\begin{equation}
    \label{eq:x_a}
    x^{C}(x-1)= c_0\,\eta \,a^{-3(1+w)(2n-1)},
\end{equation}
where $x=1+n\,\eta\,\rho^{2n-1}(1+3w^2)^{n-1}(1+3w)$, $C=(1+3w)^{-1}\left[(2n-1)(1 + 3w^2) + 8nw\right]-1$, and $c_0$ is an integration constant. The parameter $\eta$ is conveniently introduced on the RHS to ensure the correct GR limit. Equation (\ref{eq:x_a}) gives $a(\rho)$, while the inverse $\rho(a)$ can be algebraically obtained for some $C$'s. For $C=p/q$, with $p,q\in\mathbb{Z^{+}}$, $q\neq0$, restricted to $p+q\leq4$, or $C=-p/q$, with $p,q\leq4$, this can be inverted using polynomial formulas. The particularization of Eq.\ (\ref{eq:x_a}) to our cases of interest is discussed in Appendix\ \ref{sec:background_solutions}. One can also find there the formulas for $\rho(a)$ and $a(t)$ in Models A and B filled in with radiation or dust.

\section{Equations for Linear Perturbations}
In the $1+3$ covariant approach applied to Robertson-Walker space-times, developed in Refs.\ \cite{Ellis1989,Ellis1990}, cosmological perturbations are described in a geometrically transparent and gauge-invariant way by considering physically meaningful quantities projected orthogonally to the fluid four-velocity $u^\mu$. In this formalism, spatial inhomogeneities are encoded in the gradients of scalar objects such as energy density, pressure, and expansion, projected into the rest 3-space of fundamental observers comoving with the fluid.

For a general barotropic fluid, including the effective fluid arising from the $F(R,\mathcal{T})$ models discussed previously, the fundamental covariant variable representing scalar density perturbations is the comoving fractional spatial gradient of the energy density:
\begin{equation}
\label{eq:co_frac_spat_grad}
    \bar{D}_{\mu} \equiv a\,\frac{{}^{(3)}\nabla_\mu \bar{\rho}}{\bar{\rho}} ,
\end{equation}
which vanishes identically in an exact FLRW background and hence is gauge-invariant at first order. The operator ${}^{(3)}\nabla_\mu\equiv h_{\mu}{}^{\nu} \nabla_{\nu}$ denote the covariant derivative projected onto the 3-space orthogonal to $u^{\mu}$.

By considering that the effective matter content satisfies a barotropic equation of state, $\bar{p} = \bar{p}(\bar{\rho})$, the evolution of such an effective fluid is governed by the energy-momentum conservation laws, rewritten now in terms of $\bar\rho$ and $\bar p$, as follows 
\begin{align}
    \dot{\bar{\rho}} + (\bar{\rho} + \bar{p}) \Theta &= 0, \label{eq:emsg_energy_cons}\\
    (\bar{\rho} + \bar{p}) \dot{u}_\mu + {}^{(3)}\nabla_\mu \bar{p} &= 0, \label{eq:emsg_mom_cons}
\end{align}
where $\Theta = \nabla^\mu u_\mu\equiv 3H$ is the expansion coefficient, and $\dot{u}_\mu = u^\nu \nabla_\nu u_\mu$ is the 4-acceleration. The overdot here denotes \(u^b\nabla_b\). These equations are exact to the linear order and identical in structure to their GR analog, but with effective quantities $\bar\rho$ and $\bar p$.

To capture the evolution of the expansion coefficient and its spatial variation, we employ the Raychaudhuri equation:
\begin{equation}
    \dot{\Theta} + \frac{1}{3} \Theta^2 + \frac{1}{2}(\bar{\rho} + 3\bar{p}) - \Lambda - \nabla^\mu \dot{u}_\mu = 0.
\end{equation}
This equation, along with the momentum conservation law \eqref{eq:emsg_mom_cons}, plays a key role in linking the evolution of expansion gradients to that of density gradients.

\subsection{The evolution of the comoving fractional density gradient}

To derive a second-order evolution equation for $\bar{D}_{\mu}$, we follow the geometric and fully covariant procedure outlined in Refs.\ \cite{Ellis1989,Ellis1990, Bruni:1992dg, Dunsby:1991xk}, since these equations are mathematically identical to those in GR, apart from the modified sources reflecting the non-minimal coupling between matter and geometry. The method involves differentiating Eq.~\eqref{eq:emsg_energy_cons} spatially, commuting time and space derivatives, and accounting for the presence of vorticity in the fluid flow. When vorticity is present, the distribution orthogonal to $u^\mu$ is not integrable, and spatial derivatives do not commute. Specifically, the projected covariant derivatives obey a commutation relation of the form \cite{Ellis1990}
\begin{equation}
    \left[{}^{(3)}\nabla_{\mu}, {}^{(3)}\nabla_\nu\right] f = -2 \omega_{\mu\nu} \dot{f},
\end{equation}
for any scalar function $f$, where $\omega_{\mu\nu}\equiv h_\mu{}^{\alpha}h_\nu{}^{\beta}\nabla_{[\beta}u_{\alpha]}$ is the antisymmetric vorticity tensor. This non-commutativity leads to an extra source term in the second-order evolution equation for $\bar{D}_\mu$, which is absent in the irrotational case.

This term reflects a coupling between the spatial divergence of vorticity and the spatial variation of pressure, and appears explicitly in the final propagation equation. As emphasized in \ \cite{Ellis1990}, although it does not affect the clumping (scalar) part of density inhomogeneities, it contributes to dipolar and anisotropic structures, such as those relevant in Bianchi-type universes or fluids with residual turbulence.

In particular, the presence of non-zero vorticity implies that there are no spatial hypersurfaces orthogonal to the fluid flow. Consequently, the concept of comoving spatial density gradients must be defined relative to the local rest spaces of observers, and these acquire rotational contributions. This geometric insight is crucial for understanding the physical origin of the extra term and ensuring that the derivation remains fully self-consistent within the real (non-background) spacetime geometry.

The full expression for the evolution of $\bar{D}_\mu$ in a general barotropic effective fluid with nonzero vorticity is given by:
\begin{equation}
\label{eq:master_eq_bar_D}
\ddot{\bar{D}}_{\perp \mu} + \mathcal{A}(t) \dot{\bar{D}}_{\perp \mu} - \mathcal{B}(t) \bar{D}_\mu + \mathcal{L}(t)\bar{D}_\mu - \mathcal{C}(t)\, {}^{(3)}\nabla^\nu \omega_{\mu\nu} = 0,
\end{equation}
where we introduce the auxiliary coefficients 
\begin{align}
    \mathcal{A}(t)&=\left(\frac{2}{3} - 2\bar{w} + \bar{c}_s^2\right)\Theta,\label{eq:aux_A}\\
    \mathcal{B}(t)&=\left[ \left(\frac{1}{2} + 4\bar{w} - \frac{3}{2}\bar{w}^2 - 3\bar{c}_s^2\right) \bar{\rho} + (\bar{c}_s^2 - \bar{w}) \frac{12\epsilon}{a^2} + (5\bar{w} - 3\bar{c}_s^2) \Lambda \right],\label{eq:aux_B}\\
    \mathcal{C}(t)&=2 \bar{c}_s^2 a (1 + \bar{w})\, \Theta,\label{eq:aux_C}
\end{align}
and the operator 
\begin{equation}
    \mathcal{L}(t)=\bar{c}_s^2 \left( \frac{2\epsilon}{a^2}  - {}^{(3)}\nabla^2 \right).\label{eq:oper_L}
\end{equation}
The last term on the left-hand side of Eq.\ (\ref{eq:master_eq_bar_D}) arises precisely from the nonzero commutator of spatial derivatives in a rotating spacetime. The subscript $\perp$ denotes projection orthogonal to $u^\mu$. This equation generalizes the well-known evolution equation for scalar density perturbations in GR by including modified pressure and sound speed terms, as encoded in $\bar{w}$ and $\bar{c}_s^2$.

\subsection{Integrating the vorticity evolution equation}
\label{sec:vorticity}

Starting from the momentum conservation equation for the effective fluid (\ref{eq:emsg_mom_cons}) to obtain the evolution equation for vorticity, it is convenient to take the spatial curl of this equation.  Using the definition of the curl of a spatial vector as $(\operatorname{curl}X)_\mu\equiv\varepsilon_{\mu\alpha\beta}{}^{(3)}\nabla^\alpha X^\beta$, we get
\begin{equation}
\label{eq:curl_eq1}
(\bar{\rho}+\bar{p})\,\operatorname{curl}\dot u_\mu + \varepsilon_{\mu\alpha\beta}{}^{(3)}\nabla^\alpha(\bar{\rho}+\bar{p})\,\dot u^\beta+ \varepsilon_{\mu\alpha\beta}{}^{(3)}\nabla^\alpha\left({}^{(3)}\nabla^\beta \bar{p}\right) = 0.
\end{equation}
Next, we use the standard kinematic identity that relates the curl of the acceleration to the time derivative of the vorticity (valid to linear order on a Robertson–Walker background where shear and non-linear kinematic terms can be neglected). After some manipulation, Eq.\ (\ref{eq:curl_eq1}) can be written as\footnote{See Refs.\ \cite{Ellis1989,Ellis1990} for the full nonlinear identity and discussion of the shear terms which vanish for the background considered here}
\begin{equation}
\label{eq:general_vort_eq}
\dot\omega_\mu + \Big(\frac{2}{3}\Theta + \frac{\dot r}{r}\Big)\omega_\mu = 0,
\end{equation}
where we define the vorticity vector $\omega_\mu=\epsilon_{\mu\nu\alpha\beta}u^{\nu}\omega^{\alpha \beta}$ and introduce the acceleration potential \(r\) defined (up to a multiplicative constant) by
\begin{equation}
\label{eq:r_def}
r \equiv \exp\!\bigg(\int^{\bar p}\frac{d\tilde p}{\bar\rho(\tilde p)+\tilde p}\bigg),
\end{equation}
so that 
\begin{equation}
\label{eq:r_dot}
\frac{\dot r}{r} = \frac{\dot{\bar p}}{\bar\rho+\bar p}=-\bar c_s^2\,\Theta,
\end{equation} 
where we used the continuity equation and \(\bar p=\bar p(\bar\rho)\) to obtain the last equality. This equation is an exact linear relation for the evolution of the vorticity vector.

Equation~\eqref{eq:general_vort_eq} admits the immediate first integral
\begin{equation}
\label{eq:vort_first_int}
\omega_\mu(\mathbf{x},t) = \frac{\omega_\mu^{(0)}(\mathbf{x})}{a^2(t)\,r(t)},
\end{equation}
where \(\omega_\mu^{(0)}(\mathbf{x})\) is the time-independent amplitude of the vorticity mode set by the initial conditions.  Thus, aside from the universal \(a^{-2}\) dilution, the extra factor \(r^{-1}\) encodes how the time-dependence of the effective thermodynamic quantities modifies the decay (or growth) of vorticity.

Using the effective fluid description introduced in Sec.\ \ref{sec:background}, one can write \(r\) explicitly for the EMSG models considered here. A compact representation is
\begin{equation}
\label{eq:r_via_integral}
r(t) \;=\; \exp\!\left(\int^t \frac{\dot{\bar p}(\tau)}{\bar\rho(\tau)+\bar p(\tau)}\,d\tau\right)
= \exp\!\left(-\int^t \bar c_s^2(\tau)\,\Theta(\tau)\,d\tau\right),
\end{equation}
where \(\bar c_s^2\) is given by Eq.\ \eqref{eq:bar_cs2} expressed along the background solution \(\bar\rho(t)\). In important cases, this integral can be evaluated in closed form:

\paragraph{Model A with radiation.} When \(w=1/3\), the effective quantities satisfy \(\bar c_s^2=\bar w=1/3\) with \(n=1\). Hence \(\dot r/r=-\bar c_s^2\Theta=-\Theta/3\), and with \(\Theta=3H\) we obtain \(r\propto a^{-1}\). Thus
\begin{equation}
\omega_\mu \propto a^{-1}.
\end{equation}
This expression is independent of the coupling parameter $\eta$, reproducing the same behavior as GR.

\paragraph{Model A with dust.} For \(w=0\) and \(n=1\), we have
\begin{equation}
\bar c_s^2 = \frac{\eta\rho}{1+\eta\rho},\qquad \rho=\tilde\rho_0\,a^{-3},\quad \mbox{and}\quad \ r = 1 + \eta\rho,
\end{equation}
where the acceleration potential is obtained up to a constant multiplicative factor that can be absorbed in \(\omega_{\mu}^{(0)}\). 
Therefore, the first integral \eqref{eq:vort_first_int} reads in this case
\begin{equation}
\omega_\mu = \dfrac{\omega_\mu^{(0)}}{a^2\,(1+\eta\rho)}.
\end{equation}
Using \(\rho\propto a^{-3}\) this gives the explicit time dependence \(\omega_\mu\propto 1/[a^2(1+\eta\rho_0 a^{-3})]\), which interpolates between \(\omega_\mu\propto a^{-2}\) at late times (\(|\eta\rho|\ll1\)) and \(\omega_\mu\propto a\) in the very early high-density limit if \(|\eta\rho|\gg1\). It should be remarked that the strong additional suppression or enhancement will depend on the sign of \(\eta\). Situations in which the vorticity grows could play a role in the enhancement of primordial magnetic fields \cite{Marklund:2002uz, Tsagas:2001ak}.

\paragraph{Model B with any $w$.}  For the special choice \(n=\tfrac12\), the combination appearing in the effective thermodynamic variables leads to \(\bar w =\bar c_s^2\equiv\mbox{const.}\), for linear equations of state. Then, \(\dot r/r=-\bar c_s^2 \Theta\) implies \(r\propto a^{-3\bar c_s^2}\). From Eq.\ \eqref{eq:vort_first_int} we obtain the simple power-law scaling
\begin{equation}
\omega_\mu \propto a^{-(2-3\bar c_s^2)}.
\end{equation}
In particular, if \(\bar c_s^2=0\) (dust-like effective fluid) we recover \(\omega_a\propto a^{-2}\); if \(\bar c_s^2=1/3\) (radiation-like effective fluid) then \(\omega_\mu\propto a^{-1}\), decaying more slowly than in dust; and if \(\bar c_s^2>2/3\) the exponent \(2-3\bar c_s^2\) becomes negative and vorticity would grow with expansion (this indicates regimes where the effective fluid becomes very stiff and the linear analysis should be treated with caution). For comparison, when the microscopic cases of dust and radiation are considered, one obtains, respectively,
\begin{equation}
\omega^{(dust)}_\mu \propto a^{-(2-3\eta/2)},\qquad \mbox{and}\qquad \omega^{(rad)}_\mu \propto a^{(\eta\sqrt3-3)/(\eta\sqrt3+3)}.
\end{equation}
In both cases, there are critical values of $\eta$ for which the vorticity can grow with the expansion of the scale factor.

\section{Invariant Local Decomposition of \texorpdfstring{$\bar{D}_\mu$}{D\_mu}}
In order to separate the physical modes associated with the comoving fractional density gradient $\bar{D}_\mu$, we now perform a covariant and gauge-invariant local decomposition into scalar, vector, and trace-free symmetric parts. It should be remarked that such decomposition does not generate genuine vector and tensor perturbations at linear order. They are constructed from a scalar perturbation. The physical vector perturbation was previously addressed, and the shear and the magnetic part of the Weyl tensors characterize tensor perturbations. This decomposition is analogous to the standard harmonic decomposition in the metric-based approach but remains valid locally and non-perturbatively, following the formalism outlined in Ref.\ \cite{Ellis1989}.

Given that $\bar{D}_\mu$ is a spatial vector, we can consider its fully projected comoving covariant derivative $a \,{}^{(3)}\nabla_\nu \bar{D}_\mu$ and decompose it uniquely into irreducible parts:
\begin{equation}
\label{eq:irred_part_der_bar_D}
    a\,{}^{(3)}\nabla_\nu \bar{D}_\mu = \frac{1}{3}\bar\delta h_{\mu\nu} + \bar\Sigma_{\mu\nu} + \Omega_{\mu\nu},
\end{equation}
where $\bar\delta \equiv a {}^{(3)}\nabla^\alpha \bar{D}_\alpha$ is the scalar part, $\bar\Sigma_{\mu\nu} \equiv a {}^{(3)}\nabla_{(\nu} \bar{D}_{\mu)}-\frac{1}{3}\bar\delta h_{\mu\nu}$ is the symmetric and trace-free part, representing the anisotropy in the density gradient field and $\Omega_{\mu\nu} \equiv a {}^{(3)}\nabla_{[\nu} \bar{D}_{\mu]}$ is the antisymmetric part, representing the vector mode associated with the curl of $\bar{D}_\mu$. Each of these components has a clear physical interpretation: The scalar mode ($\bar\delta$) characterizes the local isotropic compression or expansion of the density field. The vector mode ($\Omega_{\mu\nu}$) arises only if the vorticity is nonzero and reflects rotational inhomogeneities. $\bar\Sigma_{\mu\nu}$ corresponds to distortion or anisotropic density gradients in the matter distribution. In particular, $\Omega_{\mu\nu}$ plays a crucial role in theories or models where vorticity does not decay rapidly, such as in some early-universe scenarios, rotating fluids, or anisotropic cosmologies. In the absence of vorticity, $\Omega_{\mu\nu}$ vanishes, and the density gradient is irrotational.

This decomposition allows us to analyze the full spatial structure of $\bar{D}_\mu$ in a covariant and gauge-invariant way, independent of any coordinate or background slicing, and serves as a basis for understanding mode coupling and the behavior of perturbations beyond the scalar sector.

\subsection{Evolution of the scalar modes}

The aforementioned scalar, vector, and tensor modes evolve according to distinct propagation equations obtained, respectively, by taking the spatial divergence, antisymmetric, and symmetric trace-free projections of the comoving covariant derivative of the master equation given by (\ref{eq:master_eq_bar_D}). 

At the linear level, only the scalar mode is relevant, in the sense that it is the only component directly linked to the clump of matter \cite{Ellis1989}. Thus, by taking the comoving divergence of Eq.\ (\ref{eq:master_eq_bar_D}) and using Eq.\ (\ref{eq:irred_part_der_bar_D}), we find:
\begin{equation}
\label{eq:scalar_mode}
    \ddot{\bar\delta} + \mathcal{A}(t) \dot{\bar\delta}
    - \mathcal{B}(t) \bar\delta- \bar{c}_s^2 {}^{(3)}\nabla^2 \bar\delta= 0.
\end{equation}
Note that the vorticity tensor does not contribute to the aggregation of matter in the linear regime.

To analyze the scale dependence of perturbations, we expand each mode in eigenfunctions $Q^{(k)}$ of the spatial Laplacian, as follows
\begin{equation}
\label{eq:scal_har_dec}
    \bar\delta = \sum_k \bar\delta^{(k)} Q^{(k)}, \quad {}^{(3)}\nabla^2 Q^{(k)} = -\frac{k^2}{a^2} Q^{(k)},
\end{equation}
where $k$ is the wavenumber. This decomposition separates the perturbations into independent scalar, vector, and tensor sectors at linear order, allowing the mode equations to be integrated independently once the background dynamics is specified.

Finally, the harmonic decomposition (\ref{eq:scal_har_dec}) of Eq.\ (\ref{eq:scalar_mode}) leads to:
\begin{equation}
\label{eq:dyn_scalar_mode}
\ddot{\bar{\delta}}^{(k)} + \mathcal{A}(t) \dot{\bar{\delta}}^{(k)}- \mathcal{B}(t) \bar{\delta}^{(k)}+ \bar{c}_s^2 \frac{k^2}{a^2}\bar{\delta}^{(k)}  = 0.
\end{equation}
Before the analysis of this equation, we shall discuss below the instability of the modes.

\subsection{Jeans instability and the time-dependent Jeans scale}
\label{sec:jeans}

Equation~\eqref{eq:dyn_scalar_mode} generalizes the standard Jeans problem to EMSG. From it, we define the instantaneous (comoving) Jeans wavenumber as \cite{EllisVanElst1999}
\begin{equation}
\label{eq:kJ_def}
k_J^2(t) \equiv \frac{a^2  \mathcal{B}(t)}{\bar c_s^2(t)} .
\end{equation}
Modes with \(k>k_J(t)\) (comoving wavelengths smaller than \(2\pi a/k_J\)) are pressure dominated and tend
to oscillate; modes with \(k<k_J(t)\) are gravitationally unstable and can grow. The novelty brought by EMSG is
encoded in the modified \(\bar c_s^2\) and \( \mathcal{B}(t)\): even for microscopic dust (\(w=0\)) one typically has
\(\bar c_s^2\neq0\) (see Eq.\ \ref{eq:dust_eff}), and thus a finite Jeans scale appears where GR would predict collapse
at all scales.

For the sake of comparison, we proceed with an approximate analytic estimate for Model A in the dust regime. For small coupling (\(|\eta\rho|\ll1\)), one may expand the effective quantities to leading order:
\begin{equation}
\bar\rho \simeq \rho + \tfrac{\eta}{2}\rho^2,\qquad \bar c_s^2 \simeq \eta\rho,\qquad \mbox{and} \qquad \bar w\simeq \tfrac{\eta\rho}{2}.
\end{equation}
Neglecting curvature and \(\Lambda\) and keeping leading terms in \(\eta\rho\), the coefficient \(\mathcal{B}(t)\) reduces to
\[
\mathcal{B}(t) \simeq \tfrac{1}{2}\bar\rho \simeq \tfrac{1}{2}\rho,
\]
so that, to leading order,
\begin{equation}
\label{eq:kJ_n1_dust}
k_J^2 \simeq \frac{a^2(\rho/2)}{\eta\rho} = \frac{a^2}{2\eta} \quad\Longrightarrow\quad
k_J \simeq \frac{a}{\sqrt{2\eta}}.
\end{equation}
Since \(k_J\) above is the comoving wavenumber, the corresponding {\em physical} Jeans length
\(\lambda_{J,\rm phys}\equiv 2\pi a/k_J\) is
\begin{equation}
\lambda_{J,\rm phys}\simeq 2\pi\sqrt{2\eta}, \qquad (\text{small }\eta\rho,\ n=1,\ \text{dust}).
\end{equation}
Hence, to leading order the physical Jeans length is approximately constant in time and determined only
by \(\eta\). This is an important qualitative result: Model A endows dust with a constant physical Jeans scale (for small \(\eta\rho\)), potentially preventing collapse below a fixed physical length even in an expanding universe. From the observational point of view, this implies that if \(\eta\) is large enough
the formation of small-scale structures could be suppressed.

On the other limit, when \(\eta\rho\gg1\), the effective quantities approach different asymptotic behavior in the Model A. We get \(\bar c_s^2\to 1\) for dust, and \(\mathcal{B}(t)\) becomes of order
\(\bar\rho\sim \eta\rho^2\). In that regime one finds \(k_J^2\sim a^2\eta\rho\) and hence the physical Jeans
length evolves nontrivially. Such high-density regimes should be treated with care because higher-order terms or microphysical considerations may be relevant.

For Model B, the algebraic form of \(\bar w\) and \(\bar c_s^2\) results in constant effective parameters for barotropic equations of state, as noted
in Sec.\ \ref{sec:flrw}. Then, from Eq.\ \eqref{eq:kJ_def}, one sees that \(k_J^2\propto a^2\bar\rho\) and therefore the physical Jeans length \(\lambda_{J,\rm phys}\sim \sqrt{\bar c_s^2/\bar\rho}\) will in general evolve with time (e.g.\ \(\lambda_{J,\rm phys}\propto a\) or
\(\lambda_{J,\rm phys}\propto a^{1/2}\) depending on the background scaling). Thus Model B typically produces a time-dependent Jeans length rather than the approximately constant length found for Model A in the small \(\eta\rho\) dust limit.

\subsection{Growth of scalar modes in Model A}
We have seen so far that EMSG models behave similarly to GR when $\eta\rho\ll1$ matching the recent past eras. In the early universe, the nontrivial coupling dominates over the other components of the Friedmann equation, particularly the spatial curvature and the cosmological constant. In this way, we shall neglect them from the further analysis, unless otherwise stated. This will simplify some equations, without significant loss of generality.

Moreover, in all numerical integrations below, we assume adiabatic initial conditions for a single fluid ($\delta\bar p=\bar c_s^2\,\delta\bar\rho$), so that no independent entropy or isocurvature mode is present. The initial amplitudes are fixed by matching the EMSG variable to its corresponding GR normalization at the initial scale factor, with the growing-mode slope used when appropriate. In a multi-fluid or non-adiabatic extension, one would have to introduce separate covariant density gradients \(D_\mu^{(i)}\), relative entropy perturbations and non-adiabatic pressure sources in the scalar propagation equations. Such an extension lies beyond the framework considered here.

\subsubsection{Radiation case} 
When $w=1/3$, the effective fluid behaves exactly as radiation, $\bar w=1/3=\bar c_s^2$, and the equation for the scalar perturbation (\ref{eq:dyn_scalar_mode}) reduces to the one from GR
\begin{equation}
\label{eq:dyn_sc_mod_A_rad}
\ddot{\bar\delta}^{(k)} + \frac{\dot a}{a} \dot{\bar\delta}^{(k)} +\left( \frac{k^{2}}{3a^2} -2\, \frac{\dot a^2}{a^2}\right)\bar\delta^{(k)}=0.
\end{equation}
Interestingly, the scale factor is $a(t)=\tilde c_0\sqrt{t}$ (given by Eq.\ \ref{eq:a_t_rad_mod_A}), even if nonlinearities are still present in the effective fluid. This equation can be fully integrated, whose solution is 
\begin{equation}
\bar{\delta}^{(k)}(t) = \left(\sqrt{\frac{\tilde t}{t}}\, c_1 -c_2\right)\cos\left(\sqrt{\frac{t}{\tilde t}}\right) +  \left(c_1 + \sqrt{\frac{\tilde t}{t}} c_2\right) \sin\left(\sqrt{\frac{t}{\tilde t}}\right),
\end{equation}
where $c_1$ and $c_2$ are integration constant, and $\tilde{t}=3\tilde{c}_0^2/4k^2$. In this case, the solution has the damping and the oscillatory regimes.

It is also instructive to consider asymptotic limits of Eq.\ \eqref{eq:dyn_sc_mod_A_rad} in terms of the wavenumber. In the super-Hubble limit $k\ll aH$, the pressure term is negligible compared with the gravitational
term and the mode equation simplifies; therefore, the perturbations evolve as
\begin{equation}
    \bar{\delta}^{(k=0)} = c_1 t + \frac{c_2}{\sqrt t}.
\end{equation}
There are growing and decaying modes similar to GR. For the sub-Hubble limit $k\gg aH$, the pressure term dominates and the modes are oscillatory with approximate frequency \(\bar c_s k/a\). The amplitude damping is controlled by \(\mathcal{A}(t)\) and the cosmological redshifting of the wave. In summary, one recovers the
standard acoustic oscillations with sound speed \(1/\sqrt3\).

\subsubsection{Dust case}
\label{scalar_dust_mod_A}

After harmonic decomposition and passing to the scale-factor variable $a$, the linear equation for the scalar comoving density-gradient perturbation (\ref{eq:dyn_scalar_mode}), in the dust-dominated epoch for Model A, takes the second-order form
\begin{equation}\label{eq:delta-general}
\bar{\delta}''_{(k)}(a) + P(a)\,\bar\delta'_{(k)}(a) + Q(a)\,\bar\delta_{(k)}(a) \;=\; 0,
\end{equation}
where primes denote derivatives with respect to $a$. The coefficient functions are rational in $a$ and depend parametrically on $(k,\eta,\alpha\equiv \eta\,c_0)$:
\begin{align}
P(a) &= 3\,\frac{a^{6}+a^{3}\alpha-\alpha^{2}}{a\,(2a^{3}+\alpha)\,(a^{3}+\alpha)}, 
\label{eq:P-def}\\[6pt]
Q(a) &= 3\,\frac{a\big(4 a^{4}\eta k^{2} - 2a^{6} + 2a\,\eta k^{2}\alpha - \alpha^{2}\big)}
{(2a^{3}+\alpha)^{2}(a^{3}+\alpha)},
\label{eq:Q-def}
\end{align}
where we used Eq.\ (\ref{eq:rho_a_Model_A_dust}) for the energy density. This equation is a linear, homogeneous, second-order ODE with four regular singular points (including infinity) and, for generic parameter values, it does not admit a closed-form solution in terms of elementary functions.

In the infinite-wavelength limit ($k=0$), the coefficient $Q$ simplifies and the resulting equation belongs to the Heun class after a simple algebraic change of variable (because of the four regular singular points at $a^{3}=0,\,-\alpha,\,-\tfrac12\alpha$, and $\infty$). Thus, the exact super-Hubble solution can be written in the form
\begin{eqnarray}
\bar\delta_{(k=0)} \left( a \right) &=& \frac {c_1}{\sqrt {2\,{a}^{3}+\alpha}}{\rm HeunG} \left( \frac{1}{2},0,-\frac{5}{6},0,-\frac{1}{3},-1,y \right)\nonumber\\ 
&&+ \frac {c_2\,{a}^{4}}{\sqrt {2\,{a}^{3} + \alpha}}{\rm HeunG} \left( \frac{1}{2},\frac{4}{3},\frac{1}{2},\frac{4}{3},\frac{7}{3},-1,y \right),\label{eq:Heun-solution}
\end{eqnarray}
with $y=-a^{3}/\alpha$. 

For large comoving wavenumber, the $k^2$-dependent terms in $Q(a)$ dominate. In this regime, it is convenient to apply the WKB approximation by putting the ODE in its normal form. By defining the integrating factor 
\begin{equation}\label{eq:mu-def}
\mu(a) \;=\; \exp\!\Big(-\tfrac12\int^a P(s)\,ds\Big)
\;=\; a^{3/2}\,(a^{3}+\alpha)^{1/2}\,(2a^{3}+\alpha)^{-5/4},
\end{equation}
with the substitution $\bar{\delta}_{(k)}(a)=\mu(a)\,\nu(a)$, Eq.\ \eqref{eq:delta-general} becomes
\begin{equation}\label{eq:y-normal}
\nu''(a) + R(a)\,\nu(a) = 0,
\end{equation}
with $R(a) = Q(a) - \tfrac12 P'(a) - \tfrac14 P(a)^2$. Then, we extract the leading large-$k$ piece of $R(a)$ by writing $R(a) = k^{2} S(a) + \widetilde R(a)$, where $S(a)$ collects the part proportional to $k^{2}$. Explicitly, it is
\begin{equation}\label{eq:S-def}
S(a) \;=\; \frac{6\eta\,a^{2}}{(2a^{3}+\alpha)(a^{3}+\alpha)}.
\end{equation}
One may check that $S(a)>0$ for physically sensible positive parameters and $a>0$, so the WKB phase is oscillatory. Therefore, for $k\gg1$, the dominant equation is $\nu'' + k^{2}S(a)\nu\approx0$, and the standard WKB leading-order solution reads
\begin{equation}\label{eq:WKB-y}
\nu(a)\;\simeq\; S(a)^{-1/4}\Big\{ A\cos\!\big(k\Phi(a)\big) + B\sin\!\big(k\Phi(a)\big)\Big\},
\qquad
\Phi(a)\equiv\int^{a}\sqrt{S(s)}\,ds.
\end{equation}
Returning to the original variable, we obtain the leading WKB approximation
\begin{equation}\label{eq:WKB-delta}
\bar\delta_{(k\gg 1)}(a)\;\simeq\;\mu(a)\,S(a)^{-1/4}\Big\{ A\cos\big(k\Phi(a)\big) + B\sin\big(k\Phi(a)\big)\Big\},
\end{equation}
with $\mu(a)$ given in \eqref{eq:mu-def} and $S(a)$ in \eqref{eq:S-def}. The WKB approximation holds provided $k$ is large enough that successive derivatives of $S(a)$ are small compared with $k^2 S(a)^{2}$. 

Setting $\eta=0$ in the coefficients \eqref{eq:P-def}--\eqref{eq:Q-def} reduces \eqref{eq:delta-general} exactly to the GR dust perturbation equation in the scale-factor variable,
\[
\delta''(a) + \frac{3}{2a}\,\delta'(a) - \frac{3}{2a^{2}}\,\delta(a) = 0,
\]
whose independent solutions are the standard power laws
\[
\delta_{\rm GR}^{(1)}(a)=a \qquad\text{(growing mode)},\qquad
\delta_{\rm GR}^{(2)}(a)=a^{-3/2}\qquad\text{(decaying mode)}.
\]
Note that in a dust-dominated universe, GR predicts a scale-invariant growth of the perturbations, contrary to Model A, which gives a faster growth of large scales but a strong damping of small scales.

\subsubsection{Back to the physical density contrast}

The analysis carried out in the previous subsections has been performed on the comoving fractional density gradient \(\bar\delta^{(k)}\), associated with the effective energy density $\bar\rho$, which obeys the modified propagation equations (\ref{eq:dyn_scalar_mode}), (\ref{eq:dyn_sc_mod_A_rad}), and (\ref{eq:delta-general}), that follow naturally from the effective-fluid rephrasing of the EMSG field equations. The physical harmonic amplitude \(\delta^{(k)}\), the one that will be responsible for \textit{matter} perturbations, is algebraically related to the effective variable by a model-dependent multiplicative mapping of the form
\begin{equation}\label{eq:delta_mapping}
\delta^{(k)} =
\frac{1+\eta\,\rho^{2n-1}A(n,w)}{1+2n\eta\,\rho^{2n-1}A(n,w)}\;\bar\delta^{(k)},
\end{equation}
where we used Eqs.\ (\ref{eq:rho_eff}), (\ref{eq:co_frac_spat_grad}), and (\ref{eq:irred_part_der_bar_D}) for the derivation. This algebraic relation is exact within the effective-fluid treatment and must be applied to any effective result (power spectra, growth factors, transfer functions) when comparing with the observable density contrast. The mapping is perturbative when \(|\eta\rho^{2n-1}A(n,w)|\ll1\), but it can induce substantial changes in the amplitude and scale-dependent corrections when the combination \(\eta\rho^{2n-1}\) is not small.

It is worth adding a few words about the meaning of the ``physical'' density contrast appearing in Eq.~(\ref{eq:delta_mapping}). The effective-fluid formulation is a convenient way of writing the modified field equations in a GR-like form, but the density contrast directly associated with matter clustering is the one constructed from the unbarred matter density \(\rho\). We therefore define
\[
D_\mu \equiv a\,\frac{{}^{(3)}\nabla_\mu \rho}{\rho}, \quad\mbox{and} \quad
\delta \equiv a\,{}^{(3)}\nabla^\mu D_\mu ,
\]
and denote its harmonic amplitude by \(\delta^{(k)}\). This is the variable to be compared with the usual matter density contrast in the GR limit. By contrast,
\[
\bar D_\mu \equiv a\,\frac{{}^{(3)}\nabla_\mu \bar\rho}{\bar\rho},
 \quad\mbox{and} \quad \bar\delta \equiv a\,{}^{(3)}\nabla^\mu \bar D_\mu ,
\]
is the corresponding quantity for the effective density \(\bar\rho\). Both \(D_\mu\) and \(\bar D_\mu\) vanish identically on the FLRW background. They are therefore first-order gauge-invariant variables. Consequently, Equation~(\ref{eq:delta_mapping}) is an algebraic relation between two gauge-invariant density perturbations: the effective density contrast used to derive the propagation equation and the physical matter density contrast used for comparison with observables.

The map (\ref{eq:delta_mapping}) leads to the following expression for Model A
\begin{equation}
\label{eq:phys-vs-eff-general}
\delta^{(k)} \;=\;
\frac{1+\tfrac12\,\eta\,\rho\,\bigl(3w^2+1+8w\bigr)}
     {1+\eta\,\rho\,\bigl(3w^2+1+8w\bigr)}\;\bar\delta^{(k)}.
\end{equation}
For any $w$, the prefactor is strictly between $1/2$ and $1$ for $\eta\rho>0$, so the mapping $\bar\delta^{(k)}\mapsto\delta^{(k)}$ produces a mild, time–dependent suppression that fades as $\rho\!\to\!0$.

\begin{figure}[t]
    \centering
    \includegraphics[width=0.49\linewidth]{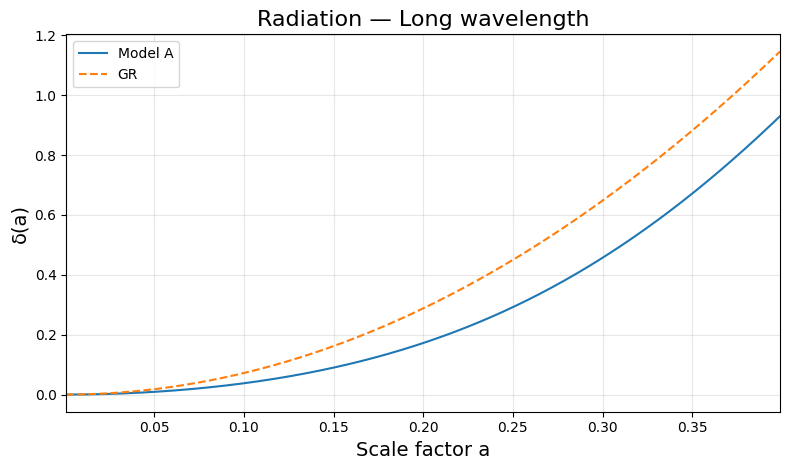}
    \includegraphics[width=0.49\linewidth]{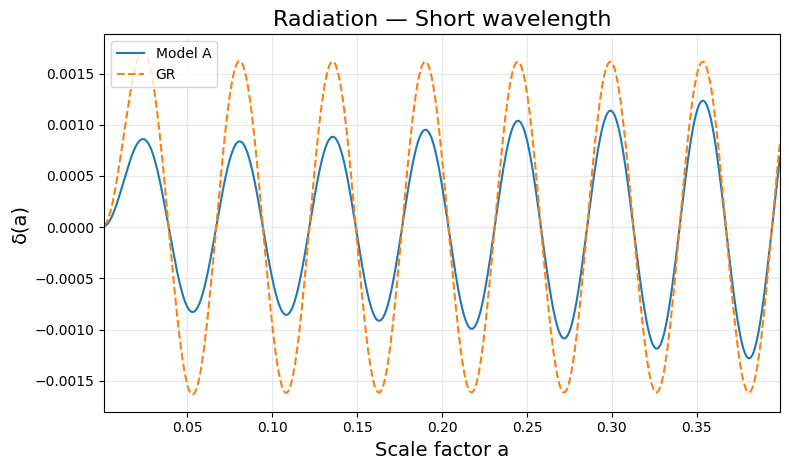}
    \caption{Model A (radiation). Physical density contrast $\delta^{(k)}(a)$ compared with GR for
    (left) $k_{\rm long}=0.01$ and (right) $k_{\rm short}=100$. Initial conditions at last scattering
    $a_\ast=1/1100$ use the same amplitude as GR, $\delta(a_\ast)=\delta_\ast=10^{-5}$, and the GR
    growing–mode slope $\delta'(a_\ast)=\delta_\ast/a_\ast$. Parameters:
    $\eta=0.01$, $\tilde c_0=1$. The final $a$ is chosen when $\delta\approx1$ (linear regime) in the long wavelength regime and also used for short wavelength, for the sake of comparison. In both panels, the Model~A curve shows a early–time suppression relative to GR as encoded in Eq.\ (\ref{eq:phys-vs-eff-rad}).}
    \label{fig:rad_model_A}
\end{figure}

In the case of radiation, Eq.~\eqref{eq:phys-vs-eff-general} becomes
\begin{equation}
\label{eq:phys-vs-eff-rad}
\delta^{(k)} \;=\; \frac{1+2\,\eta\,\rho}{1+4\,\eta\,\rho}\;\bar\delta^{(k)}\,.
\end{equation}
Fig.\ \ref{fig:rad_model_A} shows the physical contrast $\delta^{(k)}(a)$ in Model~A against the GR solution for two representative scales ($k_{\rm long}=0.01$ and $k_{\rm short}=100$). We adopt CMB–motivated
initial conditions at last scattering, $a_\ast=1/1100$, namely
$\delta(a_\ast)=\delta_\ast$ and the GR growing–mode slope $\delta'(a_\ast)=\delta_\ast/a_\ast$. In both regimes, the Model~A curve tracks the GR one but with a smaller early–time
slope, i.e., the suppression induced by the prefactor in \eqref{eq:phys-vs-eff-rad}. As $a\to1$ the curves become indistinguishable because the mapping tends to unity. Therefore, in a radiation era, Model~A yields a
slightly reduced growth of the physical density contrast at fixed $k$ relative to GR, with negligible difference at late times.

\begin{figure}[t]
    \centering
    \includegraphics[width=0.49\linewidth]{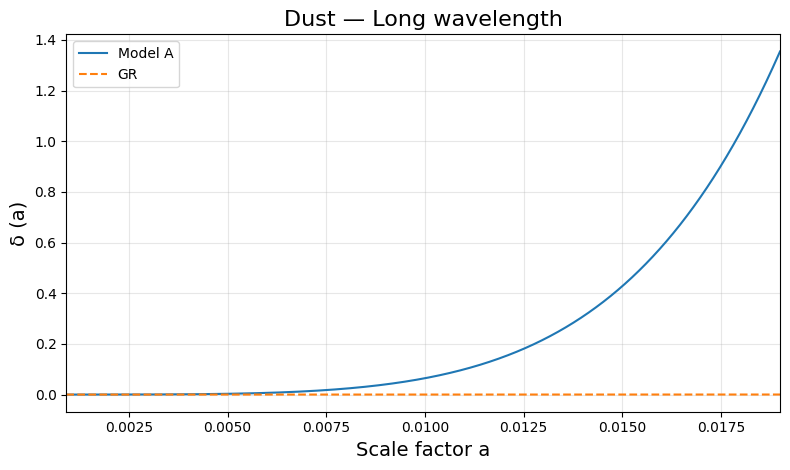}\hfill
    \includegraphics[width=0.49\linewidth]{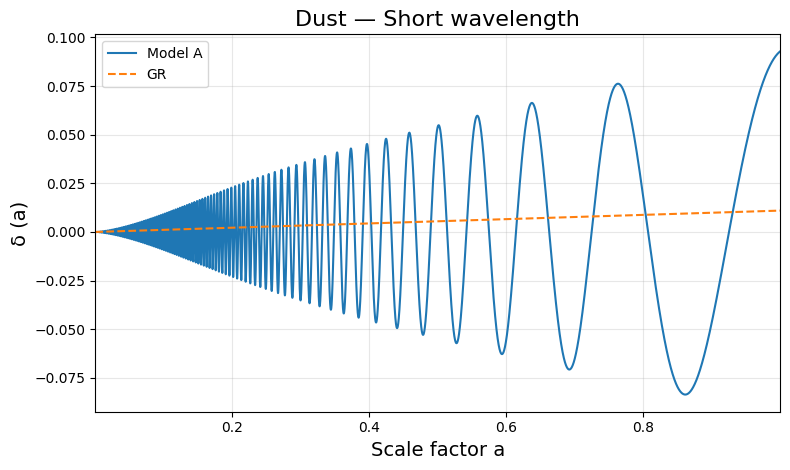}\\[4pt]
    \centering
    \includegraphics[width=0.55\linewidth]{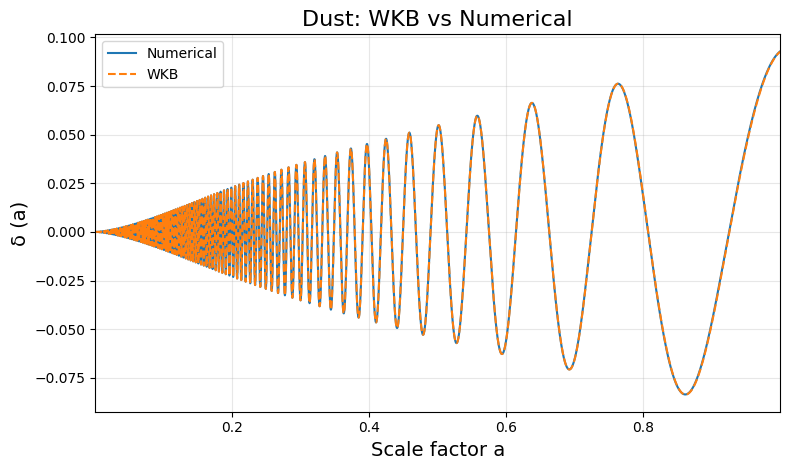}
    \caption{Model A (dust). Physical density contrast $\delta^{(k)}(a)$ compared with GR.
    \textbf{Top–left:} $k=0$, the dynamics in Model~A drives super–Hubble growth that can exceed the GR growing mode when both are initiated with the same $(\delta,\delta')$ at $a_\ast$. Again, the final $a$ is chosen to keep the linear order of perturbations in the long wavelength regime. \textbf{Top–right:} subhorizon solutions ($k=100$) show oscillations with a
    $k$–dependent envelope; the mapping suppresses the amplitude relative to the effective mode and the curves approach
    GR as $a\to1$. \textbf{Bottom:} for $k=100$, the WKB solution closely tracks the numerical physical solution. Setup: $\eta=0.01$, $c_0=10^{-3}$, $a_\ast=1/1100$, $\delta(a_\ast)=10^{-5}$, and $\delta'(a_\ast)=\delta_\ast/a_\ast$.}
    \label{fig:dust_mod_A}
\end{figure}

For dust, Eq.\ \eqref{eq:phys-vs-eff-general} gives 
\begin{equation}
\label{eq:phys-vs-eff-dust}
\delta^{(k)}(a)= \frac{1+\tfrac12\,\eta\,c_0\,a^{-3}}{1+\eta\,c_0\,a^{-3}}\;\bar\delta^{(k)}(a)\,,
\end{equation}
namely, a time–dependent suppression of the physical amplitude when $\eta c_0 a^{-3}\gtrsim1$ that smoothly approaches unity for $a\to1$. Since Eq.~\eqref{eq:delta-general} has rational coefficients, it is more convenient to solve it numerically; the physical contrast follows from
\eqref{eq:phys-vs-eff-dust}. The three panels of Fig.~\ref{fig:dust_mod_A} summarize our main findings:

(i) \emph{Long wavelength ($k=0$).} The physical mode in Model~A grows faster than the GR growing mode when initialized with the same $(\delta,\delta')$ at $a_\ast$, an effect that arises exclusively from the solution for $\bar\delta$.

(ii) \emph{Short wavelength ($k=100$).} In the subhorizon regime, the Model~A solutions display
$k$–dependent oscillations superposed on an envelope that departs from the GR growing mode.

(iii) \emph{Short–wave WKB vs numerical ($k=100$).} After casting \eqref{eq:delta-general} into Schr\"odinger form
[cf.\ Eq.~(\ref{eq:y-normal})\,], the leading–order WKB solution, given by Eqs.~(\ref{eq:WKB-y})–(\ref{eq:WKB-delta}), provides an accurate approximation to the effective mode provided the constants are fixed in an appropriate window of validity of the regime. Mapping back with \eqref{eq:phys-vs-eff-dust} yields an excellent match to the physical numerical curve.

\subsection{Growth of scalar modes in Model B}
\label{sec:IV.D.ModelB}

Again, we explore that the effective equation-of-state parameter and the effective sound speed coincide and are constant, so that the perturbation dynamics acquires an important simplification. In this model, the scalar amplitude \(\bar\delta^{(k)}(t)\) satisfies the second-order ODE
\begin{equation}\label{eq:MB_simplified}
\ddot{\bar\delta}^{(k)} + 3\Big(\tfrac{2}{3}-\bar w\Big)\frac{\dot a}{a}\dot{\bar\delta}^{(k)}
+ \Bigg[\,9\Big(\tfrac12 \bar w^{2}-\tfrac{1}{3}\bar w-\tfrac{1}{6}\Big)\frac{\dot a^{2}}{a^{2}}
+ \frac{\bar w\,k^{2}}{a^{2}}\,\Bigg] \bar\delta^{(k)} = 0.
\end{equation}

In the Appendix \ref{model_B_app}, one can find the scale factors for Model B. In both radiation and dust, the background is a power law
\begin{equation}\label{eq:a_powerlaw}
a(t) \;=\; A\,t^{p}, \qquad p \;=\; \frac{2}{3(1+\bar w)},
\end{equation}
by inserting \(\bar w =\bar w_r\) and the normalization \(A\) given by Eq.\ (\ref{a_t_mod_B_rad}) for radiation, and \(\bar w=\bar w_d\) with Eq.\ (\ref{a_t_mod_B_dust}) for dust. Substituting \(a(t)\) and its derivative into \eqref{eq:MB_simplified} yields a time-power ODE with a regular singular point at \(t=0\):
\[
\ddot{\bar\delta}^{(k)} + \frac{p\left(2 - 3\bar w\right)}{t}\dot{\bar\delta}^{(k)} + \frac{3p^{2}\left(3\bar w^{2}-2\bar w-1\right)}{2t^{2}} {\bar\delta}^{(k)} + \frac{\bar w k^{2}}{A^{2}t^{2p}} {\bar\delta}^{(k)} = 0.
\]

In the long-wavelength limit, it is convenient to seek power-law solutions \({\bar\delta}^{(0)}\propto t^{s}\), leading to a quadratic equation for \(s\). A direct substitution shows that the two roots are
\begin{equation}
\label{eq:exp_mod_B_k0}
s_1=\frac{2}{3}\frac{1+3\bar w}{1+\bar w}, \qquad
s_2=\frac{\bar w-1}{1+\bar w}.
\end{equation}
Note that $s_1$ is negative only for the interval $-1<\bar w<-1/3$, while $s_2<0$ occurs for $-1<\bar w<1$. Otherwise, both exponents are positive. With small positive \(\bar w\), the growing exponent is slightly modified from the GR value; for large \(\bar w\), the time behavior can differ qualitatively and must be treated with care due to divergences that certain choices of $\eta$ might cause. 

Now, to obtain the short-wavelength asymptotic behavior, it is useful to write \(\bar{\delta}^{(k\gg1)}(t)=\mu(t)\,\nu(t)\) with
\begin{equation}\label{eq:mu_def_MB}
\mu(t) = a(t)^{\frac{3\bar w-2}{2}}.
\end{equation}
Thus, the equation for \(\nu\) becomes
\[
\ddot \nu + R(t)\,\nu = 0,
\]
where \(R(t)\) is obtained from the algebraic combination
\[
R(t) = \Big[\,\frac{3}{2}\big(3 \bar w^{2}-2\bar w-1\big)\frac{\dot a^{2}}{a^{2}}
+ \frac{\bar w\,k^{2}}{a^{2}}\,\Big] -\left(1-\frac32\bar w \right)\left(\frac{\dot a}{a}\right)^{\cdot}
- \left(1-\frac32\bar w \right)^2\left(\frac{\dot a}{a}\right)^{2}.
\]
For large comoving wavenumber \(k\) the term proportional to \(k^2\) dominates, so to leading order
\[
R(t) \simeq k^{2}\,S(t),\qquad S(t)\simeq \frac{\bar w}{a(t)^{2}} \qquad (k\to\infty).
\]
Hence, the dominant short-wavelength equation is
\[
\ddot \nu + k^{2}\frac{\bar w}{a^{2}}\,\nu \approx 0.
\]
The standard WKB leading-order solution is then
\[
\nu(t)\simeq S(t)^{-1/4}\Big\{A\cos\!\big(k\Phi(t)\big)+B\sin\!\big(k\Phi(t)\big)\Big\},
\qquad
\Phi(t)\equiv\int^{t}\sqrt{S(\tau)}\,d\tau = \sqrt{\bar w}\int^{t}\frac{d\tau}{a(\tau)}.
\]
Returning to the original variables, we obtain the compact WKB approximation
\begin{equation}\label{eq:MB_WKB_final}
\bar{\delta}^{(k\gg1)}(t)\;\simeq\;\bar w^{-1/4}\,\mu(t)\,a(t)^{1/2}\,\Big\{A\cos\big(k\Phi(t)\big)+B\sin\big(k\Phi(t)\big)\Big\},
\qquad (k\to\infty),
\end{equation}
with \(\mu(t)\) given in \eqref{eq:mu_def_MB}. Therefore, for short wavelengths, the mode is rapidly oscillatory with phase \(k\Phi(t)\) and the slowly varying envelope scales as \(a^{(3\bar w-1)/2}\). In particular, if \(\bar w<\tfrac{1}{3}\) the envelope decays with expansion; if \(\bar w=\tfrac{1}{3}\), the envelope is constant in \(a\); and  
 if \(\bar w>\tfrac{1}{3}\) the envelope grows with \(a\).

These statements follow directly from \eqref{eq:MB_WKB_final} and provide a simple physical interpretation of the short-wavelength regime: the sign and magnitude of the effective equation-of-state parameter \(\bar w\) control whether acoustic oscillations are damped or amplified as the universe expands. Setting \(\eta\to0\) in Model B yields \(\bar w\to w\). In that limit, Eq.~\eqref{eq:MB_simplified} reduces to the usual GR perturbation equation in the chosen background, as previously discussed. 

Concerning the physical amplitude for the comoving fractional density gradient in Model B, this case displays a considerable simplification: the algebraic factor in Eq.\,\eqref{eq:delta_mapping} reduces to unity, i.e.
\begin{equation}\label{eq:delta_equality}
\delta^{(k)} = \bar\delta^{(k)}.
\end{equation}
This identity has important conceptual and practical consequences.  Conceptually, it means that the effective-fluid variables introduced in Sec.\ \ref{sec:background} are not just a calculational convenience for Model B: at linear order, the effective density gradient \(\bar\delta^{(k)}\) \emph{is} the physical (observable) density contrast.  Practically, all analytic solutions, asymptotic formulae, and numerical results derived above are directly physical predictions without any additional amplitude rescaling.

Two consequences worth emphasizing for Model B are the following.  First, since \(\bar w=\bar c_s^2\) is constant for barotropic microscopic matter in this model, the time-dependence of both the background and the perturbations takes the simple power-law/WKB form discussed above.  Therefore, the mapping between theoretical predictions and observables is immediate.  Second, because no extra algebraic factor intervenes, comparisons with GR are transparent: taking the limit \(\eta\to0\) yields \(\bar w\to w\) and the usual GR results are recovered continuously for any $k$. This makes Model B particularly convenient for illustrating the qualitative effects of a nonzero effective sound speed while avoiding the additional interpretational step required in Model A. For the sake of completeness and comparison, we discuss below the cases of radiation and dust in detail.

\begin{figure}[t]
    \centering
    \includegraphics[width=0.49\linewidth]{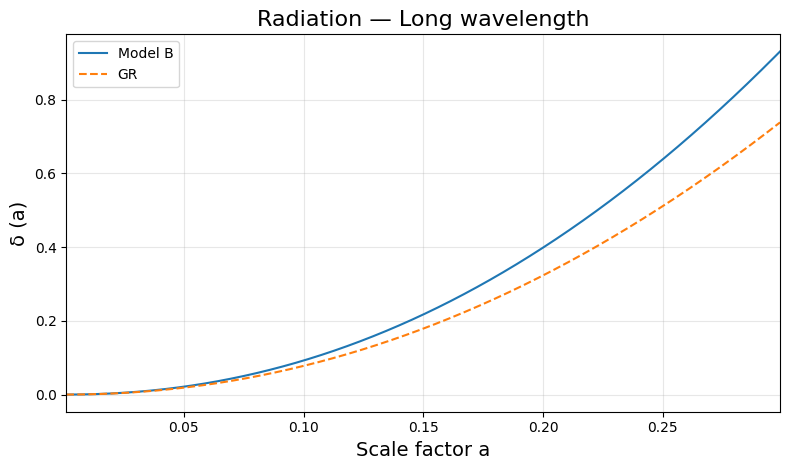}\hfill
    \includegraphics[width=0.49\linewidth]{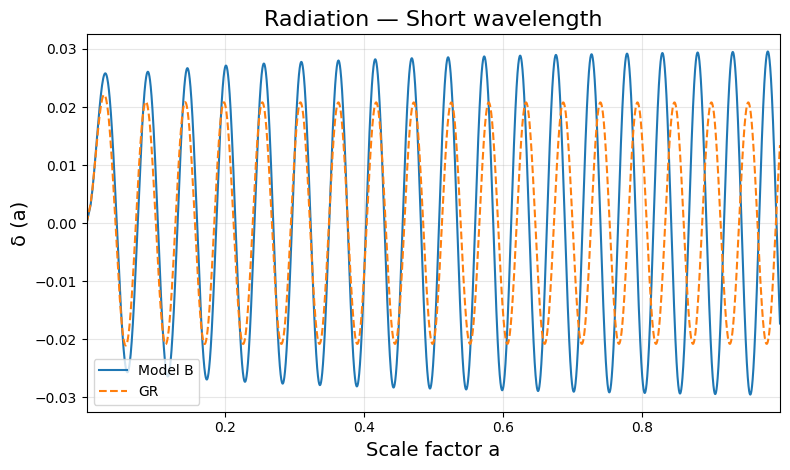}
    \caption{Model B (radiation). Physical contrast $\delta^{(k)}(a)$ compared with GR radiation. Left: long wavelength ($k=0.01$). Right: short wavelength ($k=100$). Initial conditions at last scattering $a_\ast=1/1100$ use the same amplitude as GR, $\delta(a_\ast)=\delta_\ast=10^{-5}$, and $\dot\delta(t_\ast)=0$.
    Parameters: $\eta=0.01$, $A=1$. For $\eta>0$ we have $\bar w_{\rm rad}\gtrsim 1/3$, hence a small Hubble friction and a short--wave envelope $\propto a^{(3\bar w_{\rm rad}-1)/2}$ that slowly grows, in agreement with the plotted evolution.}
    \label{fig:MB_rad}
\end{figure}

\begin{figure}[t]
    \centering
    \includegraphics[width=0.49\linewidth]{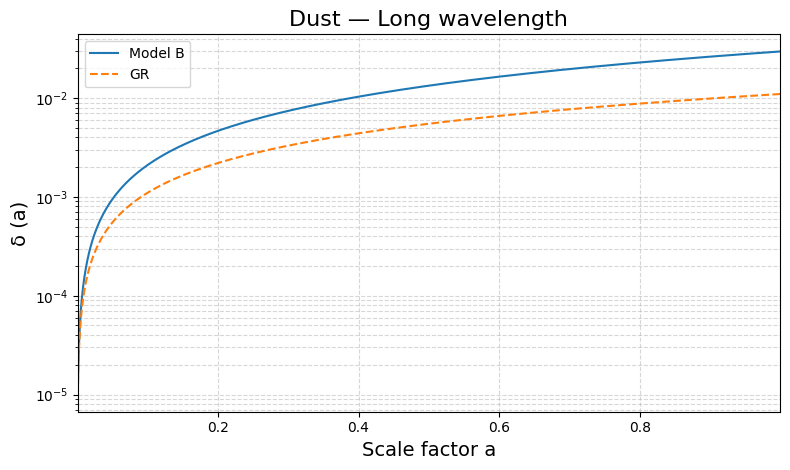}\hfill
    \includegraphics[width=0.49\linewidth]{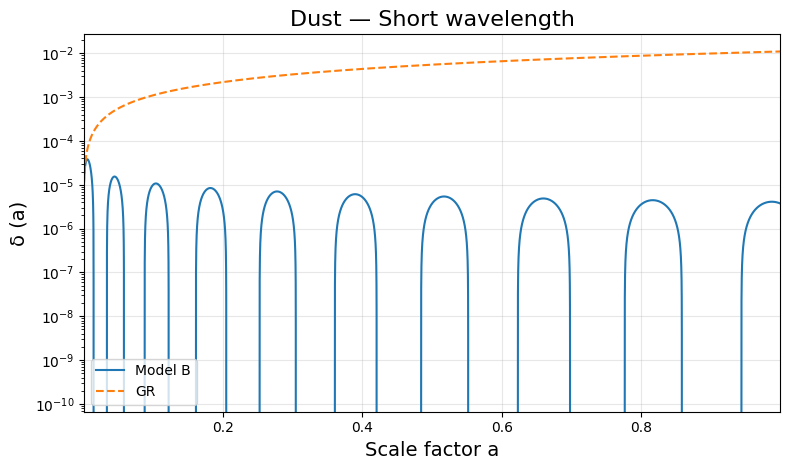}
    \caption{Model B (dust). Physical contrast $\delta^{(k)}(a)$ compared with GR radiation. Left: long wavelength  ($k=0$). Right: numerical subhorizon solutions for $k=100$ compared with the GR growing mode $\delta_{\rm GR}(a)=\delta_\ast(a/a_\ast)$. The envelope decay with $a$, while the increased Hubble friction further suppresses growth, explaining the rapid damping seen in the panel. Initial data at last scattering: $\delta(a_\ast)=10^{-5}$ and $\dot\delta(t_\ast)=p\,\delta_\ast/t_\ast$ with $p=2/[3(1+\bar w_{\rm dust})]$. Parameters: $\eta=0.1$, $A=1$; Scale factor range $0<a<1$.}
    \label{fig:MB_dust}
\end{figure}

\paragraph{Radiation case $(\bar w=\bar w_{\rm rad})$.}
For $\eta>0$ one has $\bar w_{\rm rad}\gtrsim 1/3$. Two immediate consequences follow from Eq.\ 
\eqref{eq:MB_simplified}. First, the Hubble ``friction'' term $3(2/3-\bar w)H$ is smaller than in GR, so damping is weaker. Second, the short--wave (subhorizon) frequency is $\sim \bar w\,k^2/a^2$, and the
standard WKB envelope scales as $|\delta_{\rm WKB}(t)|\ \propto\ a^{\frac{3\bar w-1}{2}}$. Because $3\bar w_{\rm rad}-1\gtrsim 0$, the envelope grows slowly with $a$. These two effects explain the numerical curves in the left/right panels of Fig.~\ref{fig:MB_rad}: long--wavelength modes track GR with a slightly larger slope, while subhorizon modes oscillate with a gently increasing envelope. The deviation from GR weakens as $a\to1$ because backgrounds converge and the overall Hubble damping fades, but the curves do not need to coincide at the
final time since their dynamics are genuinely different for $\bar w\neq 1/3$.

\paragraph{Dust case $(\bar w=\bar w_{\rm dust})$.} Here $\bar w=\eta/2$ is small and positive. Then $3\bar w-1<0$, so the short--wave WKB envelope scales with a negative exponent: subhorizon modes oscillate with a decaying envelope. Moreover, $3(2/3-\bar w)H$ is larger than in GR (stronger friction), further suppressing growth. This is precisely what Figs.\ \ref{fig:MB_dust} show: for $k=100$ the oscillations are visible, but their amplitude diminishes with expansion. On super-Hubble scales ($k=0$), the $\bar w\,k^2/a^2$ term drops out and the competition between the $H^2$-term and Hubble friction sets the slope. For the parameters shown (e.g.\ $\eta=0.1$), the $k=0$ curve grows more rapidly than the GR growing mode. For the sake of readability, we show a log–scale version of the plots.

The mode-by-mode results derived in this section should be regarded as the single-fluid building blocks for a future matter power-spectrum analysis. A genuine observable \(P(k,z)\) requires the construction of transfer functions in a multi-component cosmological model, including radiation, baryons, cold dark matter, dark energy, and possible entropy or non-adiabatic sources. Implementing the present covariant EMSG perturbation equations in such a Boltzmann-level framework lies beyond the scope of this work and will be
addressed elsewhere.

\section{Covariant approach to gravitational waves}\label{sec:tensor}

In the $1\!+\!3$ covariant approach, gravitational radiation is naturally described by the transverse–traceless tensors associated with the magnetic part of the Weyl tensor, $\mathcal{H}_{ab}$, and with the shear of the flow, $\sigma_{ab}$ \cite{Dunsby1997}. This is characterized by the conditions $\bar{D}_a=0$ and $\omega^a=0$, i.e., scalar and vector modes are zero. 
Upon harmonic decomposition on the constant–curvature hypersurfaces, each tensorial harmonic amplitude will be denoted by $\mathcal{H}^{(k)}$ and $\sigma^{(k)}$, respectively (we suppress the superscript $(k)$ when no confusion arises). Linearizing about a spatially flat FLRW background and using the background Friedmann relation to eliminate explicit densities in favor of $H\equiv \dot a/a$, the tensor sector is governed by
\begin{align}
\ddot{\mathcal{H}}+7\frac{\dot a}{a}\,\dot{\mathcal{H}}
+\left[6(1-\bar w)\left(\frac{\dot a}{a}\right)^{\!2}-\frac{k^2}{a^2}\right]\mathcal{H}&=0,
\label{eq:tensor-master-H}\\[2pt]
\ddot{\sigma}+5\frac{\dot a}{a}\,\dot \sigma
+\left[\frac{9\,\bar w+10}{2}\left(\frac{\dot a}{a}\right)^{\!2}-\frac{k^2}{a^2}\right]\sigma&=0,
\label{eq:tensor-master-sigma}
\end{align}
where $\bar w$ is the effective equation-of-state parameter specified by the model and background. Equations \eqref{eq:tensor-master-H}–\eqref{eq:tensor-master-sigma} will be our starting point; all coefficients are written in terms of $a(t)$ (or $H$), which are provided in the Appendix for each case.

For the sake of comparison and completeness, it is useful to recall \cite{Dunsby1997} the map between the covariant tensor variables used here and the metric tensor perturbations \cite{Dunsby1997}. In a spatially flat FLRW spacetime, one may write transverse-traceless components of the tensor perturbations as
\[
h_{ij}^{\rm TT} = \sum_{\lambda=+,\times} h_\lambda^{(k)}(t)\,
e_{ij}^{(\lambda)} Q^{(k)} .
\]
In an orthonormal frame, the linear shear is related to the strain $h_\lambda^{(k)}$ by
\[
\sigma_\lambda^{(k)} = \frac{1}{2}\dot h_\lambda^{(k)} ,
\]
up to a normalization factor. In its turn, the magnetic Weyl tensor is the covariant curl of the shear, so that, mode by mode,
\[
\mathcal{H}_\lambda^{(k)} \sim \frac{k}{a}\,\sigma_\lambda^{(k)}
\]
up to a polarization-dependent sign and harmonic normalization.

Therefore, the connection with the observable gravitational-wave strain is straightforward: once \(\sigma_\lambda^{(k)}\) is obtained from the covariant propagation equation, \(h_\lambda^{(k)}\) follows by time integration. Unlike the scalar case, there is no additional algebraic map analogous to Eq.~(58), because the tensor variables are purely geometrical. The EMSG parameters affect tensor observables through the background expansion and through the effective coefficients in the propagation equations, thereby modifying the tensor transfer function. The same transfer function enters the CMB tensor \(B\)-mode source term, so the \(B\)-mode signal would be shifted indirectly by the modified evolution of \(h_\lambda^{(k)}\), not by a new definition of the strain itself.

\subsection{Model A: radiation and dust backgrounds}

The background solutions $a(t)$ (and, when convenient, $\bar w$ as a function of $a$) are taken from Appendix~\ref{sec:background_solutions}. Substituting those expressions directly into \eqref{eq:tensor-master-H}–\eqref{eq:tensor-master-sigma} yields closed, density–free evolution equations for $(\mathcal H,\sigma)$.

\subsubsection{Radiation era}

For Model~A with $w=1/3$, the Appendix gives the GR–like scale factor
\begin{equation}
a(t)=\tilde c_0\,\sqrt{t},\qquad H(t)=\frac{1}{2t},
\label{eq:A-rad-at}
\end{equation}
and one has $\bar w=1/3$ exactly. Inserting these into \eqref{eq:tensor-master-H}–\eqref{eq:tensor-master-sigma} produces the explicit \emph{density–free} pair
\begin{align}
\ddot{\mathcal H}+7H\,\dot{\mathcal H}+\Big[4H^2-\frac{k^2}{a^2}\Big]\mathcal H&=0,
\label{eq:A-rad-H-Hform}\\
\ddot{\sigma}+5H\,\dot{\sigma}+\Big[\tfrac{13}{2}H^2-\frac{k^2}{a^2}\Big]\sigma&=0.
\label{eq:A-rad-sigma-Hform}
\end{align}

\paragraph*{Long wavelengths ($k=0$).}
Both equations reduce to Euler form,
$\ddot X+\frac{\alpha_X}{t}\dot X+\frac{\beta_X}{t^2}X=0$ ($X\in\{\mathcal H,\sigma\}$),
with $(\alpha_{\mathcal H},\beta_{\mathcal H})=(7/2,1)$ and $(\alpha_\sigma,\beta_\sigma)=(5/2,13/8)$. The power–law ansatz $X\propto t^{s}$ then gives $s(s-1)+\alpha_X s+\beta_X=0$, which reproduces the familiar GR radiation decay indices. Because $a(t)$ and $\bar w$ match their GR values, the long–wave tensor behavior in Model~A radiation coincides with that of GR.

\paragraph*{Short wavelengths ($k\gg aH$).}
Removing first derivatives by
\begin{equation}
\mathcal H=a^{-7/2}\nu_{\mathcal H},\qquad \sigma=a^{-5/2}\nu_\sigma,
\end{equation}
yields $\ddot\nu_X+\Omega_X^2\nu_X=0$ with
\begin{equation}
\Omega_{\mathcal H}^2=\frac{k^2}{a^2}+4H^2,\qquad
\Omega_\sigma^2=\frac{k^2}{a^2}+\frac{13}{2}H^2.
\end{equation}
For $k\gg aH$, $\Omega_X\simeq k/a$ and the standard WKB solution applies. Using $a\propto t^{1/2}$,
the phase is $\int^t\Omega_X d\tau\simeq k\int^t d\tau/a(\tau)\propto k\sqrt{t}$ and the leading envelopes follow from $\mu_X/\sqrt{\Omega_X}$:
\begin{equation}
|\mathcal H|_{\rm env}\propto a^{-3},\qquad |\sigma|_{\rm env}\propto a^{-2}\qquad (k\gg aH).
\end{equation}
Again, because $a(t)$ and $\bar w$ equal their GR counterparts, the subhorizon tensor evolution in Model~A radiation matches GR at leading order.

\subsubsection{Dust era}

For Model A dust, the Appendix provides $a(t)$ and the effective equation of state
\begin{equation}
\bar w(a)=\frac{\eta\,\rho(a)}{\eta\,\rho(a)+2},\qquad \rho(a)=c_0\,a^{-3}\,,
\label{eq:A-dust-wbar}
\end{equation}
so that $\bar w(a)>0$ and monotonically decreases to $0$ as $a\to 1$. Substituting $H=\dot a/a$ and $\bar w(a)$ into \eqref{eq:tensor-master-H}–\eqref{eq:tensor-master-sigma} gives
\begin{align}
\ddot{\mathcal H}+7H\,\dot{\mathcal H}+\Big[6\big(1-\bar w(a)\big)H^2-\frac{k^2}{a^2}\Big]\mathcal H&=0,
\label{eq:A-dust-H-Hform}\\
\ddot{\sigma}+5H\,\dot{\sigma}+\Big[\tfrac{9\,\bar w(a)+10}{2}\,H^2-\frac{k^2}{a^2}\Big]\sigma&=0.
\label{eq:A-dust-sigma-Hform}
\end{align}

\paragraph*{Long wavelengths ($k=0$).}
On an expanding dust background $H>0$ and $\bar w(a)>0$. Equation \eqref{eq:A-dust-sigma-Hform} then has a larger restoring coefficient than in GR, implying \emph{faster} shear decay. Equation \eqref{eq:A-dust-H-Hform} differs from GR only through the factor $(1-\bar w(a))$ multiplying $H^2$; since $\bar w(a)\ll1$ for most of the evolution, the magnetic Weyl amplitude follows GR closely (decaying on expansion), with small $\bar w(a)$–driven shifts in the decay index that vanish as $\bar w(a)\to0$.

\paragraph*{Short wavelengths ($k\gg aH$).}
The same redefinitions $\mathcal H=a^{-7/2}\nu_{\mathcal H}$, $\sigma=a^{-5/2}\nu_\sigma$ yield
\begin{equation}
\Omega_{\mathcal H}^2=\frac{k^2}{a^2}+6\big(1-\bar w(a)\big)H^2,\qquad
\Omega_\sigma^2=\frac{k^2}{a^2}+\frac{9\,\bar w(a)+10}{2}\,H^2.
\end{equation}
For $k\gg aH$ one again has oscillations with phase $\simeq k\int dt/a$ and leading envelopes
\begin{equation}
|\mathcal H|_{\rm env}\propto a^{-3},\qquad |\sigma|_{\rm env}\propto a^{-2},
\end{equation}
while the slow $a$–dependence of $\bar w(a)$ induces mild, model–specific tilts relative to GR. Since $\bar w(a)>0$, the $\sigma$ envelope damps more strongly than in GR; the $\mathcal H$ envelope remains very close to the GR scaling, with tiny corrections controlled by $\bar w(a)$.

\subsection{Model B: radiation and dust backgrounds}
In Model~B the effective equation of state is \emph{constant}, $\bar w=\bar c_s^{\,2}$, shifted by the EMSG coupling. The background therefore follows a power law,
\begin{equation}
a(t)=A\,t^{p},\qquad p=\frac{2}{3(1+\bar w)},\qquad H(t)=\frac{\dot a}{a}=\frac{p}{t}.
\label{eq:MB-back-Hform}
\end{equation}
We analyze the tensor sector using the density–free master equations (obtained by eliminating $\bar\rho$ in favor of $H=\dot a/a$):
\begin{align}
\ddot{\mathcal H}+7H\,\dot{\mathcal H}
+\Big[6(1-\bar w)H^{2}-\frac{k^{2}}{a^{2}}\Big]\mathcal H&=0,
\label{eq:MB-tensor-H-Hform}\\[2pt]
\ddot{\sigma}+5H\,\dot{\sigma}
+\Big[\tfrac{9\bar w+10}{2}\,H^{2}-\frac{k^{2}}{a^{2}}\Big]\sigma&=0.
\label{eq:MB-tensor-sigma-Hform}
\end{align}
All coefficients are thus expressed solely in terms of $a(t)$ and the constant $\bar w$.

In what follows we scrutinize the two physically relevant specializations, \emph{radiation–like} and \emph{dust–like}, adopting the expressions for $\bar w$ inferred in Appendix~A (also used in the scalar sector):
\begin{equation}
\bar w_{\rm rad}=\frac{\tfrac{1}{3}+\tfrac{\eta}{\sqrt{3}}}{1+\tfrac{\eta}{\sqrt{3}}}, 
\qquad 
\bar w_{\rm dust}=\frac{\eta}{2}.
\label{eq:MB-ws-const}
\end{equation}
Initial data are set at last scattering, $a_\ast=1/1100$, using the same physical amplitudes as in GR for each tensor variable; for super–Hubble radiation–like conditions we take $\dot X(t_\ast)=0$ (constant leading mode), while for the dust–like case we match the GR growing–mode slope, $\dot X(t_\ast)=p\,X(t_\ast)/t_\ast$.

\subsubsection{Radiation case}

For $\eta>0$ one has $\bar w_{\rm rad}>1/3$, hence $p=2/[3(1+\bar w_{\rm rad})]<1/2$. Substituting \eqref{eq:MB-back-Hform} into \eqref{eq:MB-tensor-H-Hform}–\eqref{eq:MB-tensor-sigma-Hform} shows that the coefficients scale as $t^{-1}$ (friction) and $t^{-2}$ (mass and wavelength terms). The long–wavelength system ($k=0$) reduces to the Euler form
\begin{equation}
\ddot X+\frac{\alpha_X}{t}\,\dot X+\frac{\beta_X}{t^{2}}\,X=0,
\qquad X\in\{\mathcal H,\sigma\},
\end{equation}
with
\begin{equation}
\alpha_{\mathcal H}=7p,\qquad \beta_{\mathcal H}=6(1-\bar w_{\rm rad})\,p^{2}, 
\qquad
\alpha_{\sigma}=5p,\qquad \beta_{\sigma}=\tfrac{9\bar w_{\rm rad}+10}{2}\,p^{2}.
\label{eq:MB-rad-betas-Hform}
\end{equation}
The power–law indices $s$ follow from $s(s-1)+\alpha_X s+\beta_X=0$. Increasing $\bar w$ relative to GR radiation ($\bar w=1/3$) decreases $\beta_{\mathcal H}$ and increases $\beta_{\sigma}$, so the magnetic Weyl amplitude $\mathcal H$ decays \emph{more slowly} at early times (delayed decay onset), whereas the shear $\sigma$ decays \emph{faster}. As the Universe dilutes, both approach the GR radiation trends but need not coincide at the final time because $(p,\bar w_{\rm rad})\neq(1/2,1/3)$.

In the subhorizon regime, $k\gg aH$, remove first derivatives via
\begin{equation}
\mathcal H=a^{-7/2}\nu_{\mathcal H},\qquad \sigma=a^{-5/2}\nu_\sigma,
\end{equation}
to obtain $\ddot\nu_X+\Omega_X^{2}\nu_X=0$ with
\begin{equation}
\Omega_{\mathcal H}^{2}=\frac{k^{2}}{a^{2}}+6(1-\bar w_{\rm rad})H^{2},
\qquad
\Omega_{\sigma}^{2}=\frac{k^{2}}{a^{2}}+\frac{9\bar w_{\rm rad}+10}{2}\,H^{2}.
\end{equation}
For $k\gg aH$ one has $\Omega_X\simeq k/a$ and the WKB form with phase $\int^{t}\Omega_X d\tau\simeq k\int^{t} d\tau/a(\tau)\propto k\,t^{1-p}$ and leading envelopes
\begin{equation}
|\mathcal H|_{\rm env}\propto a^{-3},\qquad |\sigma|_{\rm env}\propto a^{-2}\qquad (k\gg aH),
\end{equation}
while the constant shift $\bar w_{\rm rad}>1/3$ slightly weakens the early–time damping of $\mathcal H$ and strengthens that of $\sigma$ relative to GR.

\subsubsection{Dust case}

Here $\bar w_{\rm dust}=\eta/2>0$ and $p=2/[3(1+\bar w_{\rm dust})]$. The same reduction yields, for $k=0$,
\begin{equation}
\alpha_{\mathcal H}=7p,\qquad \beta_{\mathcal H}=6(1-\bar w_{\rm dust})\,p^{2},
\qquad
\alpha_{\sigma}=5p,\qquad \beta_{\sigma}=\tfrac{9\bar w_{\rm dust}+10}{2}\,p^{2}.
\label{eq:MB-dust-betas-Hform}
\end{equation}
Because $\bar w_{\rm dust}>0$ is small, $\mathcal H$ decays with an index modestly closer to zero than in GR dust (still decaying on expansion), whereas $\sigma$ decays \emph{faster} than in GR dust. In the short–wave regime, the same field redefinitions lead to
\begin{equation}
\Omega_{\mathcal H}^{2}=\frac{k^{2}}{a^{2}}+6(1-\bar w_{\rm dust})H^{2},
\qquad
\Omega_{\sigma}^{2}=\frac{k^{2}}{a^{2}}+\frac{9\bar w_{\rm dust}+10}{2}\,H^{2},
\end{equation}
and the WKB envelopes
\begin{equation}
|\mathcal H|_{\rm env}\propto a^{-3},\qquad |\sigma|_{\rm env}\propto a^{-2}\qquad (k\gg aH),
\end{equation}
with the expected $\bar w_{\rm dust}$–driven strengthening of shear damping and a very mild modification of the $\mathcal H$ envelope.

\medskip
In both radiation–like and dust–like Model~B backgrounds, expressing the tensor equations entirely in terms of $H=\dot a/a$ and the power–law index $p$ (set by the constant $\bar w$) leaves the physics transparent and density–free. Relative to GR, a larger constant $\bar w$ reduces the long–wave damping of the magnetic Weyl amplitude and increases that of the shear; subhorizon modes oscillate with phases controlled by the integral of $1/a$ and with leading envelopes $|\mathcal H|_{\rm env}\!\propto\! a^{-3}$ and $|\sigma|_{\rm env}\!\propto\! a^{-2}$. The limit $\eta\to0$ is continuous: $(p,\bar w)\to(2/3,\,w)$ and both long–wave indices and short–wave envelopes recover their GR values.

\section{Concluding remarks}\label{sec:conclusion}

We have presented a fully covariant and gauge–invariant analysis of linear cosmological perturbations in EMSG, treating on equal footing the scalar, vector (vorticity), and tensor modes within the $1\!+\!3$ formalism. A central technical simplification is that all evolution equations can be written directly in terms of the background expansion rate $H\!=\!\dot a/a$, with the scale factor $a(t)$ supplied by the Appendix, thereby avoiding explicit reference to effective densities in the perturbation dynamics. This formulation keeps the physics transparent and makes the GR limit manifest.

For barotropic, geodesic flows without anisotropic stresses, vorticity decays as
$\omega\ \propto\ a^{\,3c_s^2-2}$. Consequently, in GR radiation one has $\omega\!\propto\!a^{-1}$, while in dust $\omega\!\propto\!a^{-2}$. EMSG preserves this structural result, but modifies the decay index through the effective sound speed. In Model A dust, $\bar{c}_s^2=\eta\rho/(\eta\rho+1)$ is positive and decreases with time, implying a decay for $\omega$ that is slightly slower than $a^{-2}$ at early times and asymptotes to the GR law as $\bar{c}_s^2\!\to\!0$. In Model A radiation, $\bar c_s^2=c_s^2=1/3$, hence the GR scaling holds. In Model B, where $\bar{c}_s^2=\bar w$ is constant, the decay is a pure power law: for the radiation case, $\bar w_{\rm rad}>1/3$ gives a decay slower than $a^{-1}$, while for the dust case, $\bar w_{\rm dust}=\eta/2>0$ gives a decay slower than $a^{-2}$; both continuously recover the GR exponents as $\eta\!\to\!0$. Thus, in the absence of sources, EMSG cannot sustain growing vorticity on expanding FLRW backgrounds, but it does predict quantitatively different decay rates set by the model–dependent effective sound speed.

For scalar modes, we distinguished between the effective variable that obeys the modified propagation equation and the physical density contrast, related by an algebraic, model-dependent mapping. In Model A, the mapping is a bounded, time-dependent suppression (varying from $1/2$ at very early times to $1$ at late times), so any enhancement of growth relative to GR must originate in the effective dynamics itself. In radiation domination, Model A yields slightly smaller early–time physical amplitudes than GR and becomes indistinguishable near $a\!\to\!1$; in dust, the physical growth can exceed the GR growing mode on super–Hubble scales once the effective equation is solved and the mapping applied. In Model B, the constant effective equation-of-state parameter, $\bar w$, shifts the growth trends: $\bar w_{\rm rad}>1/3$ weakens Hubble damping in the radiation era, whereas $\bar w_{\rm dust}=\eta/2>0$ strengthens damping in a dust phase. Across regimes, short–wave solutions admit accurate WKB descriptions with phase $\propto\!\int dt/a$ and model–dependent envelopes that match the numerical analysis when constants are fixed in an adiabatic window.

In the tensor sector, the magnetic part of the  Weyl tensor $\mathcal H$ and the shear $\sigma$ are governed by two damped wave equations whose ``mass terms'' are proportional to $H^2$ with coefficients set by $\bar w$. In Model A radiation, the leading damping/envelope structure coincides with GR; the $\mathcal{O}(\eta\rho^2)$ corrections accelerate the decay of $\sigma$ at very early times and make the effective mass of $\mathcal H$ slightly more negative before diluting away. In Model A dust, the shear acquires a positive, time–dependent correction (faster decay than GR), while the magnetic Weyl envelope remains close to GR. In Model B, the constant shift of $\bar w$ cleanly organizes the phenomenology: with $\bar w_{\rm rad}>1/3$ the long–wave decay of $\mathcal H$ is reduced and that of $\sigma$ is increased relative to GR; with $\bar w_{\rm dust}>0$ both channels are more strongly damped. In all cases, subhorizon solutions oscillate with phases set by the conformal–time integral and with leading envelopes $|\mathcal H|_{\rm env}\!\propto\!a^{-3}$ and $|\sigma|_{\rm env}\!\propto\!a^{-2}$, up to controlled, slowly varying corrections. The limit $\eta\!\to\!0$ reproduces continuously the GR radiation– and dust–era results.

The restriction to radiation- and matter-dominated phases is motivated by two related points. First, the EMSG corrections considered here are controlled by powers of the matter invariant \(T_{\mu\nu}T^{\mu\nu}\), and therefore are expected to be most relevant at high densities. Second, radiation and dust provide clean barotropic backgrounds for which the covariant perturbation equations can be analyzed analytically or semi-analytically. A realistic late-time treatment would require a multi-component system including baryons, cold dark matter, radiation, and a dark-energy sector, together with the corresponding entropy and relative-density perturbations. Such an
extension is physically important but lies beyond the single-fluid adiabatic framework developed in this paper.

Concerning the two values \(n=1\) and \(n=1/2\) chosen here, they represent two qualitatively different and widely used branches of EMSG. The case \(n=1\) corresponds to the standard quadratic correction in \(T_{\mu\nu}T^{\mu\nu}\), producing density-dependent effective pressure and sound speed. The case \(n=1/2\), by contrast, leads to constant effective equation-of-state and sound-speed parameters for barotropic matter, making it an analytically tractable benchmark. These two cases therefore capture complementary behaviors of the more general power-law family.

The covariant framework developed here consolidates the linear phenomenology of EMSG and highlights robust, testable trends across sectors: early–time scalar growth tilts, tensor damping patterns, and modified vorticity decay indices. These features can be confronted with cosmological observations through (i) the shape and normalization of matter power spectra and growth indices, (ii) CMB temperature/polarization transfer functions (including integrated Sachs–Wolfe and early–ISW contributions), (iii) bounds on the stochastic gravitational–wave background and B–modes, and (iv) constraints on primordial or late–time vector perturbations via their imprints on polarization and lensing. Extending the present analysis to mixed radiation–matter eras, non-adiabatic fluids, neutrino anisotropic stress, and non–flat backgrounds will sharpen these signatures. An analysis of the growth function $f$ and the growth index $\gamma$ is currently underway to evaluate competitive constraints on the EMSG coupling and to assess possible degeneracies with standard late–time physics.

\bmhead{Acknowledgements}

PKSD is supported by a grant from the First Rand Bank (SA) and a UCT Research Committee Block Grant. EB is partially supported by CNPq (grant N.\ 305217/2022-4) and FAPEMIG (APQ-05207-23).

\begin{appendices}

\section{Background solutions for Models A and B}\label{sec:background_solutions}

In this section, we present the explicit integrated forms \(\rho(a)\) and \(a(t)\) used in the perturbation analysis. The starting point is the algebraic integration of the continuity equation given by Eq.\ (\ref{eq:x_a}). After that, we particularize it for the two representative cases studied here (Models A and B), and then we solve the resulting Friedmann equation by quadrature.

\subsection{Model A}

By substituting $n=1$ in the general effective-fluid definitions (\ref{eq:rho_eff}) and (\ref{eq:p_eff}), we obtain
\[
\bar\rho = \rho + \frac{\eta}{2}(1+8w+3w^2)\rho^2, \qquad
\bar p = p + \frac{\eta}{2}\big(1+3w^2\big)\rho^2.
\]
In this case, the auxiliary variable $x$ and the exponent $C$ simplify, and the polynomial equation (\ref{eq:x_a}) for $\rho$ reduces to
\begin{equation}
\label{eq:int_cont_mod_A}
 \left[(3\,w+1)\eta\,\rho\,+1 \right]^{{\frac {w \left( 3\,w+5 \right) }{1+3\,w}}}\eta\,\rho\, \left( 1+3\,w \right) =c_{{0}}\eta\,{a}^{-3(1+w)}.
\end{equation}
To find $\rho(a)$ for the cases of interest we have to fix the parameter $w$. Below we give the two specialized \(w\)-cases used in the paper.

\paragraph{(i) Radiation:} By setting \(w=1/3\) in the equation above, we get
\[
2\, \left( 2\,\eta\,\rho+1 \right) \rho={\frac {c_{{0}}}{{a}^{4}}}.
\]
The explicit solution becomes
\begin{equation}
\label{eq:rho_a_n1_rad}
\rho(a) \;=\; \frac{-1 + \sqrt{\,1 + 4\,\eta\,c_0\,a^{-4}\,}}{4\eta},
\end{equation}
where we already selected the branch that gives GR in the limit $\eta\rightarrow0$.

The effective energy density and pressure are
\begin{align}
\bar\rho(a) = \rho + 2\eta\rho^2,\quad \mbox{and} \quad
\bar p(a)  = \tfrac{1}{3}\rho + \tfrac{2}{3}\eta\rho^2.
\end{align}
Note that \(\bar w=\bar p/\bar\rho=1/3\) exactly for all \(\eta\) (the \(\eta\)-dependence cancels). The same happens when we combine the Friedmann (\ref{eq:friedmann_general}) and the acceleration (\ref{eq:acc_general}) equations to obtain $a(t)$ in this case. The result is the same as GR, namely
\begin{equation}
\label{eq:a_t_rad_mod_A}
a(t) = \tilde c_0\sqrt{t},
\end{equation}
where we have used the time translation freedom to eliminate spurious integration constants and defined $\tilde c_0=\sqrt{2\sqrt{c_0/6}}$. Remarkably, the evolution driven by Model A predicts a singular flat universe without a cosmological constant.

\paragraph{(ii) Dust:} Substituting \(w=0\) in Eq.\ (\ref{eq:int_cont_mod_A}), it reduces drastically, yielding
\begin{equation}
\label{eq:rho_a_Model_A_dust}
\rho=\frac{c_{0}}{a^{3}},
\end{equation}
which is the same relation as in GR.

The effective energy density and pressure are now
\begin{align}
\bar\rho(a) = \rho + \frac{\eta}{2}\rho^2,\quad \mbox{and} \quad \bar p(a)  = \tfrac{\eta}{2}\rho^2.
\end{align}
Note that \(\bar w= \eta\rho/(\eta\rho+2)\) and $\bar c_s^2=\eta\rho/(\eta\rho+1)$. 

Therefore, the Friedmann equation can be integrated and the scale factor is
\begin{equation}
\label{eq:a_t_dust_mod_A}
a(t) = \frac{6^{\frac{1}{3}}}{2}\left(t^2c_0 - 4c_1t + 4c_2\right)^{\frac{1}{3}},
\end{equation}
where the constants are subjected to the constraint $\eta c_0^2-6c_1^2+6c_2c_0=0$.
Now one can see that Model A can admit bouncing universes depending on the choice of constants.

\subsection{Model B}\label{model_B_app}

For \(n=\tfrac12\) the general effective expressions (\ref{eq:rho_eff}) and (\ref{eq:p_eff}) reduce to
\begin{align}
\bar\rho &= \rho + 2\eta\,\frac{\rho p}{\sqrt{\rho^2+3p^2}}, \label{eq:barrho_nhalf}\\
\bar p   &= p + \tfrac{\eta}{2}\sqrt{\rho^2+3p^2}, \label{eq:barp_nhalf}
\end{align}
leading to the identification of now constant parameters $\bar w = c_s^2$. A crucial simplification is that the microscopic matter conservation equation \(\dot\rho+3H(1+\bar w)\rho=0\)
still holds, but in terms of $\bar w$. Hence, the microscopic density scales in the standard way:
\begin{equation}
\rho(a) = \rho_0\,\Big(\dfrac{a_0}{a}\Big)^{3(1+\bar w)}. \label{eq:rho_a_nhalf}
\end{equation}
Thus for Model B one does not need to solve a nonlinear algebraic equation for \(\rho(a)\); instead \(\rho(a)\)
is the usual power law and the EMSG corrections enter through \(\bar\rho(a)\) and \(\bar p(a)\) given
by \eqref{eq:barrho_nhalf}--\eqref{eq:barp_nhalf}. The Friedmann equation is therefore
\begin{equation}
H^2(a) = \frac{1}{3}\bar\rho(a) = \frac{1}{3}\Big[\rho(a) + 2\eta\,w\,\frac{\rho(a)}{\sqrt{1+3w^2}}\Big],
\label{eq:H2_nhalf}
\end{equation}
and the cosmic time integral is
\begin{equation}
t - t_0 \;=\; \int_{a_0}^a \frac{d\tilde a}{\tilde a\,\sqrt{\tfrac{1}{3}\bar\rho(\tilde a)}} .
\end{equation}
Below we present the two physically relevant specializations.

\paragraph{(i) Radiation:} For the equation of state \(p=\rho/3\), we have
\begin{align}
\bar\rho & = \rho\Big(1 + \frac{\eta}{\sqrt{3}}\Big),\\
\bar p   &= \rho\Big(\frac{1}{3} + \frac{\eta}{\sqrt{3}}\Big).
\end{align}
Hence the effective equation of state is
\[
\bar w_r = \frac{\tfrac{1}{3} + \tfrac{\eta}{\sqrt{3}}}{1 + \tfrac{\eta}{\sqrt{3}}}.
\]
By using Eq.\ (\ref{eq:rho_a_nhalf}) into Eq.\ (\ref{eq:H2_nhalf}), one gets
\begin{equation}
\label{a_t_mod_B_rad}
\frac{a(t)}{a_0} = \left[\frac{3(1+\bar w_r)}{2}\sqrt{\frac{\rho_0}{3}\left(1+\frac{\eta}{\sqrt{3}}\right)}\,t\right]^{\frac{2}{3(1+\bar w_r)}}.
\end{equation}
Thus, \(a(t)\) evolves more slowly than GR if $\eta>0$, since the exponent will be smaller than $1/2$. 

\paragraph{(ii) Dust:} For dust \(p=0\), Eqs.\
\eqref{eq:barrho_nhalf}--\eqref{eq:barp_nhalf} give
\begin{align}
\bar\rho &= \rho, \label{eq:barrho_nhalf_dust}\\
\bar p   &= \frac{\eta}{2}\rho,\label{eq:barp_nhalf_dust}
\end{align}
and, therefore, $\bar w_d= \eta/2$. Again, the integration of the Friedmann equation is straightforward, yielding
\begin{equation}
\label{a_t_mod_B_dust}
\frac{a(t)}{a_0} = \left[\frac{3(1+\bar w_d)}{2}\sqrt{\frac{\rho_0}{3}\left(1+\frac{\eta}{\sqrt{3}}\right)}\,t\right]^{\frac{2}{3(1+\bar w_d)}}.
\end{equation}
Again, the power law evolution in time will succeed with an exponent smaller than that given by GR under the same circumstances when $\eta$ is positive.

\end{appendices}

\nocite*{}
\bibliography{ref}

\end{document}